\documentclass[onecolumn]{aastex631}
\usepackage{amsmath}

\usepackage{hyperref}

\newcommand{\pfrac}[2]{\left( \frac{#1}{#2} \right)}
\newcommand{\MBH}{M_{\bullet}}

\newcommand{\be}{\begin{equation}}
\newcommand{\ee}{\end{equation}}

\defcitealias{Linial&Metzger23}{LM23}

\shorttitle{QPE Disks}
\shortauthors{I.~Linial, B.D.~Metzger}

\begin{document}

\title{Coupled Disk-Star Evolution in Galactic Nuclei and the Lifetimes of QPE Sources}

\author[0000-0002-8304-1988]{Itai Linial}
\affil{Department of Physics and Columbia Astrophysics Laboratory, Columbia University, New York, NY 10027, USA}
\affil{Institute for Advanced Study, 1 Einstein Drive, Princeton, NJ 08540, USA}

\author[0000-0002-4670-7509]{Brian D.~Metzger}
\affil{Department of Physics and Columbia Astrophysics Laboratory, Columbia University, New York, NY 10027, USA}
\affil{Center for Computational Astrophysics, Flatiron Institute, 162 5th Ave, New York, NY 10010, USA} 

\begin{abstract}
A modest fraction of the stars in galactic nuclei fed towards the central supermassive black hole (SMBH) approach on low-eccentricity orbits driven by gravitational-wave radiation (extreme mass ratio inspiral, EMRI). In the likely event that a gaseous accretion disk is created in the nucleus during this slow inspiral (e.g., via an independent tidal-disruption event; TDE), star-disk collisions generate regular short-lived flares consistent with the observed quasi-periodic eruption (QPE) sources. We present a model for the coupled star-disk evolution which self-consistently accounts for mass and thermal energy injected into the disk from stellar collisions and associated mass ablation. For weak collision/ablation heating, the disk is thermally-unstable and undergoes limit-cycle oscillations which modulate its properties and lead to accretion-powered outbursts on timescales of years to decades, with a time-averaged accretion rate $\sim 0.1 \dot{M}_{\rm Edd}$. Stronger collision/ablation heating acts to stabilize the disk, enabling roughly steady accretion at the EMRI-stripping rate. In either case, the stellar destruction time through ablation, and hence the maximum QPE lifetime, is $\sim 10^{2}-10^{3}$ yr, far longer than fall-back accretion after a TDE. The quiescent accretion disks in QPE sources may at the present epoch be self-sustaining and fed primarily by EMRI ablation. Indeed, the observed range of secular variability broadly match those predicted for collision-fed disks. Changes in the QPE recurrence pattern following such outbursts, similar to that observed in GSN 069, could arise from temporary misalignment between the EMRI-fed disk and the SMBH equatorial plane as the former regrows its mass after a state transition.
\end{abstract}


\section{Introduction}

Supermassive black holes (SMBH; \citealt{Salpeter64}) with millions to billions of times the mass of our Sun, are ubiquitous in the centers of massive galaxies (e.g., \citealt{Kormendy&Richstone95}). The majority of SMBH growth over cosmic time results from radiatively-efficient accretion \citep{Soltan82}, fueled by gas funneled towards the nucleus via large-scale galactic processes (e.g., \citealt{Hopkins+16}). Nevertheless, SMBHs can grow via other mechanisms than those responsible for large-scale active-galactic nuclei (AGN). Most massive black holes are embedded in dense nuclear star clusters (e.g., \citealt{Bahcall&Wolf76,Gallego-Cano+18}) of comparable or greater mass, the constituents of which provide additional, albeit ``sequestered'', sources of mass. Stellar capture or tidal disruption events (TDE; \citealt{Hills75,Rees88}) offer another SMBH fuel source, one which may even dominate the growth of low-mass black holes (e.g., \citealt{Milosavljevic+06,Stone+17}). Most TDEs involve the destruction of main-sequence stars and are accompanied by months-long UV/X-ray emission (e.g., \citealt{Gezari+06}; though most of the total radiated energy is emitted over years to decades, e.g., \citealt{vanVelzen+19,Jonker+20}).  The tidal disruption of giant stars can generate flares lasting up to several decades (e.g., \citealt{MacLeod+12}), and may preferentially occur through multiple partial disruptions (e.g., \citealt{MacLeod+13}), potentially setting a minimum floor on the accretion rates in galactic nuclei (e.g., \citealt{MacLeod+13}; however, see \citealt{Bonnerot+16b}).

While the vast majority of TDE victim stars approach the SMBH on nearly parabolic orbits from parsec scale distances, a small fraction $\sim 1-10\%$ are predicted to reach the tidal radius due to gravitational-wave radiation on less eccentric or even quasi-circular orbits. Such stellar ``extreme mass-ratio inspirals'' (EMRIs; e.g., \citealt{Linial&Sari17,Linial&Sari23}) have been hypothesized to produce electromagnetic phenomena as rich or richer than ordinary TDEs, if and when they begin Roche-lobe overflow onto the SMBH (e.g., \citealt{Dai&Blandford13, Linial&Sari17,Metzger&Stone17,King20,King22,Metzger+22,Krolik&Linial22,Linial&Sari23,Lu&Quataert23}). However, their survival down to such small radial scales $\simeq r_{\rm T}$ is not assured.  In particular, the timescale for a stellar EMRI to inspiral through gravitational-waves across the range of radii just outside the tidal radius $r_{\rm T}$ (say, from $4 r_{\rm T}$ to $2r_{\rm T}$) is typically millions of year, far in excess of the typical interval between (ordinary) TDEs in galactic nuclei, of about once per $10^{4}-10^5$ yr (e.g., \citealt{Stone&Metzger16}). This suggests that the EMRI will inevitably interact with the transient gaseous disk formed from a TDE (\citealt{Linial&Metzger23}; hereafter 
\citetalias{Linial&Metzger23}). The generally inclined orbit of the EMRI results in the star periodically passing through the disk midplane, where it collides with dense gas and has material stripped from its surface (e.g., \citealt{Armitage+96}). These interactions can potentially lead to changes in the star's orbit (e.g., \citealt{MacLeod&Lin20,Generozov&Perets23,Linial&Quataert24}) and/or eventually result in the star's destruction (\citetalias{Linial&Metzger23}). 

Repeated star-disk collisions may also be observable directly. ``Quasi-periodic Eruptions'' (QPEs; e.g., \citealt{Miniutti+19,Giustini+20,Arcodia+21,Chakraborty+21,Arcodia+22,Miniutti+23,Webbe&Young23,Arcodia+24}) are a new class of periodic flaring X-ray sources from low-mass galactic nuclei (e.g., \citealt{Wevers+22}),
with recurrence times ranging from a few to tens of hours. The short ($\lesssim $ hours-long) eruptions from X-ray QPE systems are characterized by peak luminosities $\sim 10^{42}$ erg s$^{-1}$, typically at least comparable to those of the spectrally softer quiescent luminosity observed between eruptions \citep{Miniutti+19,Giustini+20,Arcodia+21,Chakraborty+21,Arcodia+22,Miniutti+23,Webbe&Young23}.   
While many models have been proposed for QPEs \citep{Zalamea+10,King20,Ingram+21,Xian+21,Metzger+22,Pan+22,Zhao+22,King22,Krolik&Linial22,Linial&Sari17,Kaur+23,Linial&Metzger23,Franchini+23,Tagawa&Haiman23}, only a few account for an important clue: a regular alternating behavior, observed in at least two of the QPE sources$-$GSN 069 \citep{Miniutti+19,Miniutti+23} and eRO-QPE2 \citep{Arcodia+21}, in which the temporal spacing between consecutive eruptions varies back and forth by around 10\%. 
 
\citetalias{Linial&Metzger23} showed that many of the observed properties of X-ray QPEs can be reproduced in a scenario in which a stellar EMRI collides twice per orbit with a gaseous accretion disk (see also \citealt{Sukova+21,Xian+21,Tagawa&Haiman23,Franchini+23,Zhao+22}).
The eruptions are powered by hot shocked disk material which expands from either side of midplane, akin to dual miniature supernova explosions (e.g., \citealt{Ivanov+98}). The oscillating long-short recurrence time behavior follows naturally from the longer time the star spends between collisions on the apocenter side of the disk. The timing of the eruptions is related to the schedule of disk passages, which is subject to the precession of the stellar orbit and the disk, as well as delayed caused by finite light-travel time \citep[e.g.,][]{Xian+21,Franchini+23,Chakraborty+24,Pasham+24a,Zhou+24}.

One of the outstanding questions in the disk-star-collision scenario for QPEs is the origin of the accretion flow responsible for the ``quiescent'' soft X-rays seen between eruptions. In systems where it is sufficiently luminous to detect, the quiescent spectra is consistent with thermal emission from the inner regions (few $R_{\rm g}$ scale) of a radiatively efficient accretion flow onto the SMBH at a sizable fraction $\gtrsim 0.1$ of the Eddington rate \citep{Miniutti+19,Arcodia+21,Arcodia+22,Miniutti+23,Arcodia+24}.

Long-term evolution of the quiescent luminosity (e.g., over timescales much longer than the QPE period) offers clues to the origin of the gaseous disks in these systems. While the quiescent X-ray flux of the QPE source RXJ 1301.9+2747 \citep{Giustini+20} has been remarkably constant over 20 years of monitoring, the quiescent luminosities of other QPE sources appear more variable. Eight years prior to the discovery of QPE activity, the well-studied source GSN 069 exhibited starting in 2010 an X-ray outburst similar to a TDE \citep{Shu+18,Sheng+21}, though exhibiting a longer duration and slower post-maximum decay than most TDE flares. However, starting at the end of 2019, and after QPE flares were detected in the system, it exhibited a second X-ray outburst qualitatively similar to the 2010 brightening, which was interpreted by \citet{Miniutti+23b} as being another stellar-disruption event in the same galactic nucleus, or the partial TDE of the same progenitor that gave rise to the 2010 outburst.   
\citet{Quintin_23} describe the rise of a QPE-like eruption (very similar to those seen from eRO-QPE1) in XMM-Newton, during a gradual decay in X-ray quiescence, several months following its detection as an optical TDE. \cite{Chakraborty+21} discuss a QPE candidate XMMSL1 J024916.6-041244 that demonstrated a gradual decay in X-ray flux reminiscent of GSN 069, with QPE variability seen in 2006, but none observed in 2021. Of the two new QPE sources recently detected in eROSITA by \citet{Arcodia+24}, the quiescent flux of eRO-QPE3 declined by an order of magnitude over a time span of just a few years, while in eRO-QPE4 it showed evidence of {\it brightening} (insofar that the source was not detected in archival ROSAT or during earlier eROSITA survey epochs taken just a few years prior to the first detection).   

Tentative evidence also exists for some connection between the quiescent accretion disk and QPE properties. After a year of QPE activity, the eruptions from GSN 069 become undetected at the onset of the second quiescent-emission outburst in 2019; however, as the outburst faded, the periodic eruptions recently reappeared \citep{Miniutti+23b}, albeit initially with different recurrence properties than before (G.~Minutti, private communication). A secular decrease in the average eruption luminosity from eRO-QPE3 coincided with a gradual dimming of its quiescent luminosity \citep{Arcodia+24}. \cite{Chakraborty+24} and \citet{Pasham+24a} reported that the eruptions from eRO-QPE1 dimmed by a factor of $\sim 10$ over a three year baseline, possibly with no significant variations in quiescent emission over this duration. 

While most stellar EMRIs will experience a TDE during their gravitational-wave inspiral \citepalias{Linial&Metzger23}, the complex and diverse evolution of QPE quiescent emission$-$including both decaying and rising light curves as well as re-brightenings$-$may call into question whether it is fall-back accretion from the TDE that continues to feed the SMBH and set the quiescent emission {\it at the present epoch}. Indeed, a second gas source available to feed the disk is \textit{mass stripped from the same stellar EMRI} responsible for powering the eruptions. The mass flux generated from star-disk collisions, for disk and orbital properties necessary to explain the observed eruptions, was estimated to be comparable to the black hole accretion rate implied by the quiescent emission (\citetalias{Linial&Metzger23}). While a transient TDE-disk still provides an appealing mechanism to initiate the overall process, it is natural to ask whether a disk-star collision system can becoming self-perpetuating even after TDE fall-back has abated. This longer-lived system would terminate only after the star is destroyed or if powerful collisions cease, e.g., as a result of changes in the inclination between the disk and stellar orbit.  \citet{Lu&Quataert23} similarly proposed gradual stellar mass-loss as the source of quiescent accretion in QPE systems, albeit through a different mechanism of Roche-lobe overflow regulated by the effects of ram pressure on the stellar atmosphere from disk collisions; however, such collisions will strip mass from the star, even for stellar orbits outside the tidal radius. 

In this paper, we develop a model for the coupled long-term evolution of mutually-inclined star-disk systems, which accounts for mass and thermal energy added to the disk via collisions and stellar stripping.  In Sec.~\ref{sec:model} we describe the model, in which the star is assumed to reside on a quasi-circular orbit and the disk with which it interacts is modeled as a single-annulus centered around the collision radius. In Sec.~\ref{sec:results} we present our numerical results, most of which can be understood analytically. Though simplified in many respects, the model reveals a surprisingly rich behavior, ranging from limit-cycles driven by the thermal instabilities of radiation-dominated disks, to approximately steady accretion when heating from heating from stellar-collisions/ablation is sufficiently powerful to stabilize the disk. In Sec.~\ref{sec:discussion} we discuss implications of our results for the emission properties of observed QPE sources and their demographics in galactic nuclei. In Sec.~\ref{sec:conclusions} we summarize our findings and conclude.
\section{Model}
\label{sec:model}

A stellar EMRI of mass $M_{\star} = \mathcal{M}_{\star}M_{\odot}$ and radius $R_{\star} = \mathcal{R}_{\star}R_{\odot}$ orbits the SMBH of mass $M_{\bullet}$ on a quasi-circular orbit of semi-major axis $r_0 \sim 100 R_g$ and period $P_{\rm orb} \simeq 2\pi (r_0^{3}/GM_{\bullet})^{1/2}$, where $R_{\rm g} \equiv GM_{\bullet}/c^{2}$ is the gravitational radius.  
Collisions between the star and gaseous disk happen twice per orbit, such that the average interval between observed eruptions (the QPE period) is given by $P_{\rm QPE} = P_{\rm orb}/2$ (neglecting orbital eccentricity\footnote{Throughout most of the paper we consider small eccentricities, as is expected for stellar-EMRIs, and as is interpreted from the sample of known QPEs (e.g., \citealt{Metzger+22,Linial&Sari23,Lu&Quataert23,Linial&Metzger23}; however, see \citealt{King20}). Other repeating nuclear transients, such as ASASSN-14ko \citep{Payne+21}, may also involve star-disk interaction, albeit on highly eccentric orbits \citep[e.g.,][]{Linial&Quataert24}.}).  The collision radius can thus be written:
\be
r_0 \approx 1.4 \times 10^{13}\,\,{\rm cm} \; \mathcal{P}_{\rm QPE,4}^{2/3}M_{\bullet,6}^{1/3} \approx 95 \, R_{\rm g} \, \pfrac{\mathcal{P_{\rm QPE,4}}}{M_{\bullet,6}}^{2/3} \,,
\ee
where $\mathcal{P}_{\rm QPE,4} \equiv P_{\rm QPE}/(4\,{\rm hr})$ and $M_{\bullet,6} \equiv M_{\bullet}/(10^{6}M_{\odot})$. The condition that the star does not overflow its Roche-lobe, $r_0 \gtrsim 2 r_{\rm T} \simeq 2 R_{\star}(M_{\bullet}/M_{\star})^{1/3}$, defines a minimum QPE period,
\be
P_{\rm QPE,min} \simeq \pi \left(\frac{8R_{\star}^{3}}{GM_{\star}}\right)^{1/2} \simeq 3.9\,{\rm hr}\,\mathcal{R}_{\star}^{3/2}\mathcal{M}_{\star}^{-1/2}.
\label{eq:PQPEmin}
\ee
We focus on tight orbits with $P_{\rm orb} \gtrsim P_{\rm QPE,min}$, as these stars will interact with the gaseous disks created by tidal disruption events (\citetalias{Linial&Metzger23}), the latter expected to form near the circularization radius $\simeq 2r_{\rm T}$ (for properties of the disrupted star broadly similar to the EMRI).

The star's orbital angular momentum vector makes an angle $\chi \in [0,\pi/2]$ with the spin-axis of the black hole (a retrograde orbit is not expected after some period of time, for reasons discussed below), about which it undergoes nodal precession from the Lense-Thirring torque on a timescale (e.g., \citetalias{Linial&Metzger23}):
\be
\tau_{\rm prec} \simeq \frac{P_{\rm orb}}{2a_{\bullet}}\left(\frac{r_0}{R_{\rm g}}\right)^{3/2} \simeq 1.4\,{\rm yr}\,\frac{\mathcal{P}_{\rm QPE,4}^{2}}{M_{\bullet,6}}\left(\frac{a_\bullet}{0.3}\right)^{-1},
\label{eq:Tprec}
\ee
where $a_{\bullet} \le 1$ is the dimensionless spin parameter of the black hole.  We neglect changes to the inclination angle or semi-major axis of the star's orbit, as we will show that for tight orbits, consistent with those needed to explain observed QPE eruptions, the star is destroyed by collisions faster than it can undergo appreciable radial migration or have its orbit ground into the disk midplane (Sec.~\ref{sec:orbit}). 

\begin{figure*}
    \centering
    \includegraphics[width=0.49\textwidth]
    {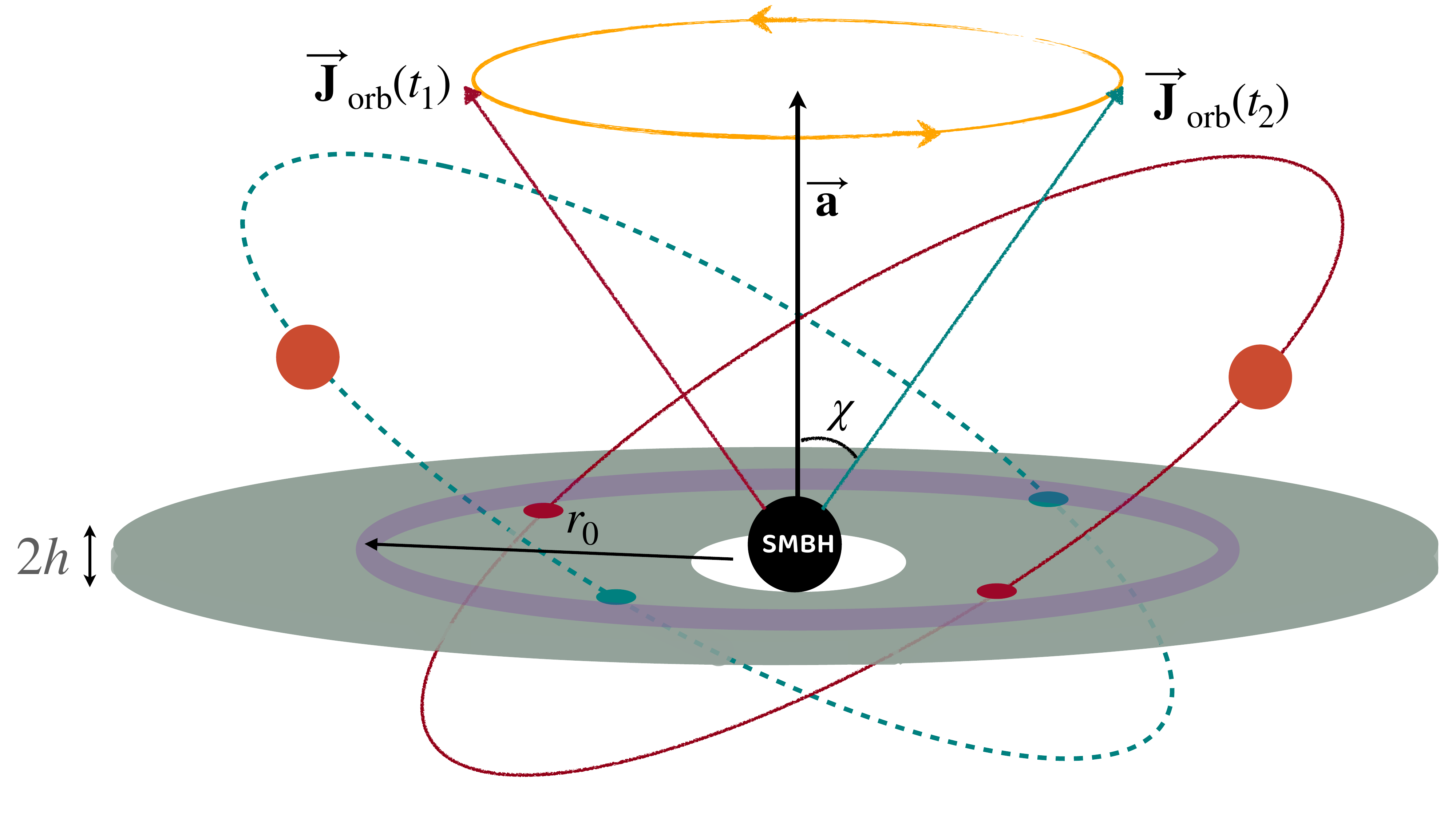}   
    \includegraphics[width=0.3\textwidth]{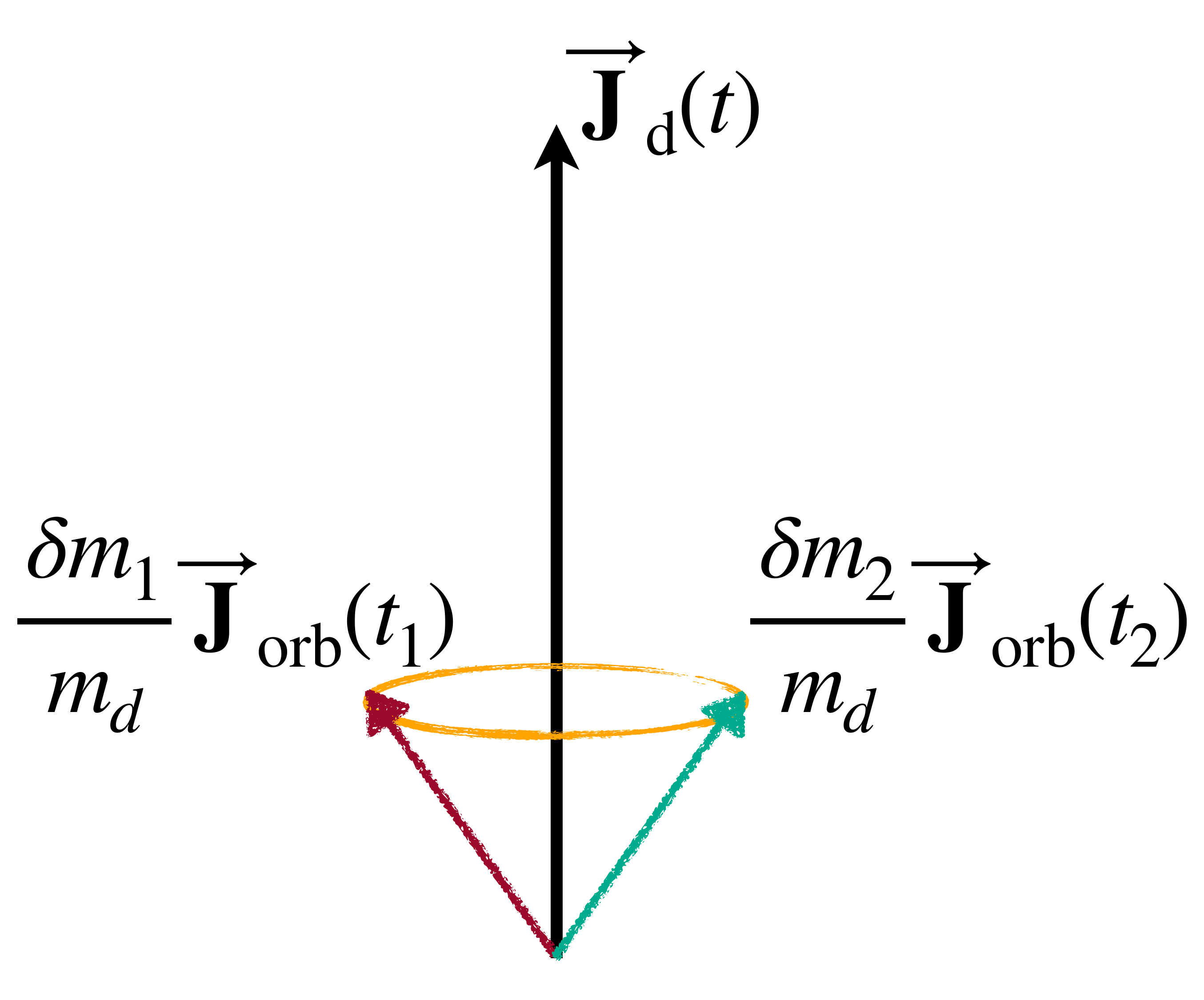}
    \caption{Schematic diagram showing the star-disk system and the long-term effects of Lense-Thirring precession on the stellar orbit. As long as the orbital precession period is rapid compared to the timescale over which the disk's surface density evolves, the net angular momentum added to the disk by the mass stripped from the star  (averaged over many precession cycles) will keep the disk aligned with the black hole spin axis. }
    \label{fig:schematic}
\end{figure*}

The angular momentum axis of the gaseous disk is assumed to align with the black hole spin axis. Other than greatly simplifying the physical picture, this assumed alignment is justified by two arguments: (1) the \citet{Bardeen&Petterson75} effect, through which dissipation of precession-driven warps in the disk eventually leads to spin-alignment out to hundreds of gravitational radii, depending on the strength of stresses within the disk (e.g., \citealt{Nelson&Papaloizou00,Sorathia+13,Zanazzi&Lai19}); (2) any component of angular momentum perpendicular to the black hole spin axis which results from mass added to the disk by stellar stripping will nearly cancel out over many stellar nodal precession cycles: we shall show that the precession time (Eq.~\eqref{eq:Tprec}) is generally shorter than the timescale over which the disk's surface density evolves (see Fig.~\ref{fig:schematic}). We expand on this justification later (Sec.~\ref{sec:orbit}), and possible observational consequences when this assumption is violated on the QPE recurrence pattern.

We model the portion of the gaseous disk with which the star interacts as an annulus of radius $r_0$ and radial extent $\sim r_0$, with a local surface density $\Sigma$, local mass $M_{\rm d} = \pi r_0^{2} \Sigma$, midplane density $\rho = \Sigma/h$, midplane temperature $T$, vertical scale-height $h \ll r_0$ and orbital speed
\be
v_{\rm orb} \simeq v_{\rm k} = \left(\frac{GM_{\bullet}}{r_0}\right)^{1/2} \approx 0.1 \, {\rm c}\;\mathcal{P}_{\rm QPE,4}^{-1/3}M_{\bullet,6}^{1/3}.
\ee 
The star loses mass $\Delta M_{\star}$ per collision with the disk, corresponding to a time-averaged mass-loss rate
\be
\dot{M}_{\star} = -\frac{\Delta M_{\star}}{P_{\rm QPE}}.
\label{eq:dMsdt}
\ee
Motivated by hydrodynamical simulations of supersonic flows past stars \citep{Armitage+96,Liu+15}, we assume that the mass stripped from the star per midplane passage is proportional to the ram pressure $p_{\rm ram} = \rho v_{\rm rel}^{2}/2$, viz.~
\be
\Delta M_{\star} = \eta \frac{p_{\rm ram}}{p_{\star}},
\label{eq:Mdotstar}
\ee
where $p_{\star} = GM_{\star}^{2}/(4\pi R_{\star}^{4})$ is the interior pressure of the star and $\eta \sim 10^{-3}$ (\citealt{Armitage+96,Liu+15}, Yao \& Quataert, in prep.).
The collision speed, corresponding to the relative motion of the star's orbit and the rotating disk, can be written $v_{\rm rel} = \xi_{\rm v} v_{\rm k}$, where $\xi_{\rm v} \in [0,2]$ is a dimensionless number that depends on the star/disk inclination angle $\chi$. For a quasi-circular orbit of eccentricity $e \ll 1$

\begin{equation}
    \xi_{\rm v} \simeq 2\sin{(\chi/2)} \left( 1+ \frac{3}{4} e \cos{f} \right) + \mathcal{O}(e^{2}) \,,
\end{equation}
where $f$ is collision's site true anomaly, measured with respect to the orbit's argument of periapsis.

Mass-loss from the star occurs sufficiently rapidly relative to its thermal timescale (typically millions of years), that the star will evolve adiabatically \citep{Linial&Sari17}. For a low-mass star of initial radius/mass $R_{\star,0}/M_{\star,0}$ whose interior is dominated by gas pressure, with an effective adiabatic index $\gamma_{\rm ad} = 5/3$, the radius increases upon mass-loss according to:
\be
R_{\star} = R_{\star,0}\left(\frac{M_{\star}}{M_{\star,0}}\right)^{-1/3}.
\label{eq:Rstar}
\ee
This expansion of the star will accelerate the mass-loss rate near the end of its life and potentially induce Roche-lobe overflow onto the black hole, somewhat hastening its final destruction.

While mass is added to the disk by star-collisions, mass is lost due to accretion onto the black hole. This occurs at the rate
\be
\dot{M}_{\rm acc} = \frac{M_{\rm d}}{t_{\rm \nu}} = 3\pi \nu_0 \Sigma,
\label{eq:Mdotacc}
\ee
where
\be
\tau_{\nu} = \frac{1}{3}\frac{r_0^{2}}{\nu_0} = \frac{1}{3}\frac{r_0}{\alpha v_{\rm k}}\left(\frac{h}{r}\right)^{-2} 
\label{eq:tnu}
\ee
is the local viscous timescale at radius $r_0$ and we have modeled the effective viscosity using the standard \citet{Shakura&Sunyaev73} prescription $\nu_0 = \alpha r_0 v_{\rm k}(h/r_0)^{2},$ where $\alpha = 0.01-0.1$ (e.g., \citealt{King+07,Davis+10}).
Even if the gaseous disk is initially formed at radii $\sim r_0$, it will spread to radii larger than $r_0$ due to outward angular momentum redistribution; it might therefore seem questionable that one can approximate the evolution of the local surface density at $r_0$ using a simple one-zone model.  However, by solving the standard time-dependent diffusion equation for $\Sigma(r,t)$ \citep[e.g.,][]{Pringle81}, with a mass-source term localized around a fixed radius ($r_0$, in our case due to star-disk collisions) one can show that as long as the source term evolves slowly compared to the local viscous timescale $\tau_{\nu}$, most of the injected mass at $r_0$ indeed flows to small radii $< r_0$, rather than accompanying the angular momentum carried to large radii (see \citealt{Metzger+12b}, their Appendix B). 

In summary, the disk mass near the collision radius evolves according to:
\be
\frac{dM_{\rm d}}{dt} = |\dot{M}_{\star}| - \dot{M}_{\rm acc} + \dot{M}_{\rm TDE}.
\label{eq:dMddt}
\ee
The final term accounts for an external mass source, which is needed to start the collision-fed disk evolution. The primary example here is fall-back from the eccentric debris streams of a tidally-disrupted star, which for a full-disruption obeys (e.g., \citealt{Phinney89})
\be
\dot{M}_{\rm TDE} = \frac{M_{\rm \star,TDE}}{3t_{\rm TDE}}\left(\frac{t}{t_{\rm TDE}}\right)^{-5/3},
\label{eq:MdotTDE}
\ee
where $t > t_{\rm TDE}$ is measured since the time of disruption, $M_{\rm \star,TDE}$ is the mass of the disrupted star (distinct from the still-present quasi-circular EMRI) and $t_{\rm TDE}$ is the fall-back time of the most tightly bound debris (typically weeks to months for main-sequence stars; e.g., \citealt{Stone+13}).

Each midplane passage, the star shocks a quantity of disk material corresponding to that intercepted by its physical cross section:
\be \Delta M_{\rm coll} \simeq 2 \pi R_{\star}^{2}\Sigma. 
\label{eq:Mcoll}
\ee  
where the factor of 2 accounts for the increase in swept-up material mass for $\chi\approx \pi/4$ (\citetalias{Linial&Metzger23}). Although this hot fast expanding material may be key to generating QPE eruptions as it rises above the disk midplane and radiates away its energy (\citetalias{Linial&Metzger23}; Sec.~\ref{sec:flares}), the velocity dispersion it acquires from the collision $\lesssim v_{\rm k}$ is not sufficient to travel appreciably away from the collision radius $\sim r_0$. We therefore neglect any such mass-loss term in Eq.~\eqref{eq:dMddt} insofar that we are modeling an annulus of radial width $\sim r_0$ centered at $r_0.$ Even were this not the case, in practice we typically find that the mass shocked and ejected from the disk by each collision is much smaller than that stripped from the star, i.e. $\Delta M_{\rm coll} \ll \Delta M_\star$.

The midplane pressure of the disk includes contributions from gas and radiation, 
\be
p = p_{\rm gas} + p_{\rm rad} = \frac{\rho k_{\rm B}T}{\mu m_p} + \frac{a}{3}T^{4},
\label{eq:EOS}
\ee
where $\mu$ is the mean molecular weight (we take $\mu = 0.62$ for fully-ionized solar-composition gas). Vertical hydrostatic balance gives an expression,
\be
\left(\frac{h}{r}\right)^{2} - \frac{1}{3}\frac{r_0}{\Sigma v_{\rm k}^{2}}aT^{4} \left(\frac{h}{r}\right) - \frac{k_{\rm B}T}{\mu m_p v_{\rm k}^{2}} = 0,
\label{eq:hoverreq}
\ee
for the disk aspect ratio in terms of the midplane temperature $T$, yielding
\begin{equation} \label{eq:hovereq_gen}
    \frac{h}{r} = \frac{a T^4 r_0}{6 \Sigma v_{\rm k}^2} \left( 1 + \sqrt{1 + \frac{k_{\rm B} T}{\mu m_p} \pfrac{6\Sigma v_{\rm k}}{a T^4 r_0}^2 } \right) \,.
\end{equation}

A final equation follows the evolution of the disk's thermal energy $E_{\rm th} = \left[(3/2)p_{\rm gas} + 3p_{\rm rad}\right]2\pi h r_0^{2}$, 
\be
\frac{dE_{\rm th}}{dt} = \dot{E}_+ - \dot{E}_-,
\label{eq:dEthdt}
\ee
as a result of various sources of heating $\dot{E}_+$ and cooling $\dot{E}_-$.  Heating results from both viscous dissipation (``accretion'') and energy deposited by the stellar collisions and ablation,
\be
\dot{E}_+ = \dot{E}_{\rm acc} +\dot{E}_{\rm abl} = \left( \frac{3}{8} \dot{M}_{\rm acc} + \beta \left(|\dot{M}_{\star}| + \dot{M}_{\rm coll}\right) \right) v_{\rm k}^2 ,
\ee
where $\dot{M}_{\rm coll} = \Delta M_{\rm coll}/P_{\rm QPE}$ is the time-averaged rate of mass excavation from the disk due to collisions. In the collision/ablation heating term, we assume that a fixed fraction of the specific kinetic energy $\sim v_{\rm k}^{2}$ of the matter stripped ($|\dot{M}_{\star}|$) or intercepted ($\dot{M}_{\rm coll}$) by the star's passage through the disk goes into heating the midplane, where the efficiency of this process is encapsulated in the dimensionless parameter $\beta \lesssim 1$. 

For short orbital periods characteristic of most QPE sources we shall find that $\dot{M}_{\rm coll} \ll |\dot{M}_{\star}|$, i.e. the rate that mass is stripped from the star far exceeds the rate at which the star intercepts disk material (this condition is roughly equivalent to the star's destruction occurring prior to its 
orbit undergoing significant drag-induced orbital evolution). While the detailed evolution of the disk will turn out to be sensitive to $\beta$, its value is uncertain because it depends on where and how efficiently the kinetic energy of the ablated stellar mass is incorporated into the disk midplane prior to being radiated. We note that there is no a-priori reason that ablation and collision would heat the disk with the same coupling coefficient $\beta$. However, since typically $\dot{M}_{\rm coll} \ll |\dot{M}_{\star}|$ in the regimes of interest, this simplification is not expected to have any significant impact on our results. 

Cooling occurs via vertical radiative diffusion and radial advection, viz.
\be
\dot{E}_{-} = \dot{E}_{\rm rad} + \dot{E}_{\rm adv} = \frac{4}{3}\pi r_0^{2}\frac{\sigma T^{4}}{\tau} + |\dot{M}_{\rm acc}|\frac{P}{\rho}\zeta,
\label{eq:therm1}
\ee
where $\tau = \kappa \Sigma$ is the optical depth through the disk midplane, $P$ is the disk's midplane pressure and the dimensionless parameter $\zeta \propto d{\rm ln} s/d{\rm ln} r$ depends on the radial entropy gradient and is typically of order unity (e.g., \citealt{DiMatteo+02}). We take $\zeta = 4$, though our results are not sensitive to this choice because the disk typically spends little if any time in the geometrically-thick state where advective cooling is important. For the opacity law, 
\be
\kappa = \kappa_{\rm T} + \kappa_{\rm K},
\label{eq:opacity}
\ee
we include contributions from electron scattering $\kappa_{\rm T} \approx 0.34$ cm$^{2}$ g$^{-1}$ and Kramers' opacity $\kappa_{\rm K} = A_{\rm \kappa} \rho T^{-7/2}$, as appropriate for temperatures $T \gtrsim 10^{4}$ K, where $A_{\kappa} \approx 5\times 10^{24} \; \rm cm^5 \, gr^{-2} \, K^{7/2}$.

In thermal equilibrium ($\dot{E}_{+} = \dot{E}_-$), Eq.~\eqref{eq:dEthdt} becomes
\be
\beta v_{\rm k}^{2}\left(|\dot{M}_{\star}| + \dot{M}_{\rm coll}\right) + \frac{9\pi}{8}\alpha \Sigma r_0 v_{\rm k}^{3}\left(\frac{h}{r}\right)^{2} = \frac{4}{3}\pi r_0^{2}\frac{\sigma T^{4}}{\Sigma \kappa} 
 + 3\pi \zeta \alpha \Sigma r_0 v_{\rm k}^{3}\left(\frac{h}{r}\right)^{4}.
\label{eq:therm2} 
 \ee
 
To summarize the model, we solve Eqs.~\eqref{eq:dMsdt}, \eqref{eq:dMddt}, \eqref{eq:dEthdt}, supplemented by Eqs.~\eqref{eq:hoverreq}, \eqref{eq:opacity}. We resolve the disk's thermal time (hence, allow for $\dot{E}_{\rm th} \neq 0$), but assume that hydrostatic equilibrium is always maintained (as enforced by Eq.~\eqref{eq:hoverreq}). A model is fully specified by the parameters: $\{M_{\bullet},M_{\star,0}, R_{\star,0},\alpha,\chi,\eta, \beta,\mathcal{P}_{\rm QPE} \gtrsim \mathcal{P}_{\rm QPE,min} \}$ in addition to those parameters  $\{ M_{\rm \star,TDE}, t_{\rm TDE}\}$ which characterize the external source-term $\dot{M}_{\rm TDE}(t)$.  Table \ref{tab:modelparams} summarizes the model variables and the fiducial values adopted in the numerical models we present. 

\begin{deluxetable}{lll}
\tablecaption{Model Parameters\label{tab:modelparams}}
\tablewidth{700pt}
\tabletypesize{\scriptsize}
\tablehead{
\colhead{Symbol} & \colhead{Description} & 
\colhead{Fiducial Value(s)} 
} 
\startdata
$M_{\bullet}$ & SMBH mass & $10^{6}M_{\odot}$ \\
$M_{\star,0} = \mathcal{M}_{\star}M_{\odot}$ & Initial EMRI star mass & $1M_{\odot}$ \\
$R_{\star,0} = \mathcal{R}_{\star}R_{\odot}$ & Initial EMRI star radius & $1R_{\odot}$ \\
$P_{\rm QPE} = P_{\rm orb}/2$ & QPE period & 4 hr \\
$\chi$ & Star-disk inclination  & $\pi/4$ \\
$\beta \propto \dot{E}_{\rm abl}/v_{\rm k}^{2}$ & Collision/ablation heating efficiency & 0...0.25 \\
$\alpha$ & Disk viscosity parameter & 0.1 \\
$\eta \propto \Delta M_{\star}/p_{\rm ram}$ & Collision stripping efficiency & $10^{-3}$ \\
$M_{\rm \star,TDE}$ & Mass of disrupted star in TDE & $1M_{\odot}$\\
$t_{\rm TDE}$ & Fallback time of TDE debris & 30 d\\
\enddata
\end{deluxetable}

\section{Results}
\label{sec:results}

We present results for the coupled disk/star system evolution for an initial EMRI similar to the Sun ($M_{\star,0} = M_{\odot};$ $R_{\star,0} = R_{\odot}$) on an orbit of half-period $P_{\rm QPE} = 4$ hr characteristic of QPE sources. Assuming that a steady-state can be achieved in which mass accretes inwards through the disk at the same rate it is stripped from the star (i.e., $\dot{M}_{\rm acc} \approx \dot{M}_{\star}$), the ratio of collision/ablation- and accretion-heating rates (left two terms in Eq.~\eqref{eq:therm2}) takes the form:
\be
\frac{\dot{E}_{\rm abl}}{\dot{E}_{\rm acc}} \approx \frac{8}{3}\beta.
\ee
We shall find that the system can exhibit qualitatively different evolution depending on this ratio, i.e. on the value of $\beta$ (even though some solutions only achieve the steady-state $\dot{M}_{\rm acc} \approx \dot{M}_{\star}$ in a time-averaged sense). 

We first consider the limit of negligible collision/ablation heating ($\beta = 0$; Sec.~\ref{sec:beta0}) before considering models with progressively stronger collision/ablation heating (larger $\beta$; Secs.~\ref{sec:midbeta}). While for low values of $\beta$ the disk is thermally unstable and exhibits limit-cycle behavior, we find that for sufficiently large $\beta > \beta_{\rm stable} = 3/16$ collision/ablation heating can act to stabilize the disk evolution (Sec.~\ref{sec:stable}). Finally, we consider a star on a long-period orbit, for which the disk is gas-pressure dominated and hence thermally stable even for $\beta = 0$ (Sec.~\ref{sec:longperiod}). The dependence of these different regimes on $\beta$ and $P_{\rm QPE}$ is discussed in \ref{sec:orbit}.

\subsection{Thermally Unstable Evolution}
\subsubsection{Weak collision/ablation heating}

\label{sec:beta0}

In a standard steady-state accretion flow model (e.g., neglecting collision/ablation heating), radiation-pressure dominates over gas-pressure in the disk midplane at small radii (e.g., \citealt{Linial&Metzger24}),
\be
r_0 \lesssim r_{\rm rad} \approx 5.4\times 10^{13}\,{\rm cm}\, \alpha_{-1}^{2/21}M_{\bullet,6}^{23/21}\left(\frac{\dot{M}_{\rm acc}}{0.1\dot{M}_{\rm Edd}}\right)^{16/21},
\ee
or, correspondingly short QPE periods,
\be
P_{\rm QPE} \lesssim 30\,{\rm hr}\, \alpha_{-1}^{1/7}M_{\bullet,6}^{8/7}\left(\frac{\dot{M}_{\rm acc}}{0.1\dot{M}_{\rm Edd}}\right)^{8/7},
\label{eq:radpressure}
\ee
where we have normalized the accretion rate to the Eddington value $\dot{M}_{\rm Edd} \equiv L_{\rm Edd}/(\epsilon c^{2})$, $L_{\rm Edd} \simeq 1.5\times 10^{44}M_{\bullet,6}$ erg s$^{-1}$ and $\epsilon = 0.1$. As most QPE sources have periods $\sim 2-20$ hr and $\dot{M} \gtrsim 0.1\dot{M}_{\rm Edd}$ (as inferred from their quiescent emission), radiation pressure can be important in the disk near the collision annulus.

For $r_{0} < r_{\rm rad}$ and $\beta = 0$ (i.e., the typical case absent stellar collisions), Eqs.~\eqref{eq:hoverreq}, \eqref{eq:therm2} are known to admit a range of optically-thick disk solutions, depending on the surface density $\Sigma$ (e.g., \citealt{Chen+95}). For low $\Sigma$ below a critical threshold value $\Sigma_{\rm min}$ the only solutions are radiatively-cooled ($\dot{E}_+ \simeq \dot{E}_{\rm rad}),$ and geometrically-thin $h/r \ll 1.$  The cooler solution is gas pressure-dominated and thermally-stable, while the hotter solution is radiation pressure-dominated and thermally unstable (e.g., \citealt{Pringle&Rees72,Hirose+09,Jiang+19}). On the other hand, for very large $\Sigma$ above another critical value $\Sigma_{\rm max}$ the only solution is advectively cooled ($\dot{E}_+ \simeq \dot{E}_{\rm adv}),$ radiation-dominated, and geometrically-thick $h/r \sim 1$, i.e. even hotter than the thermally-unstable radiatively-cooled branch (e.g., \citealt{Abramowicz+88}).  At intermediate $\Sigma_{\rm min} < \Sigma < \Sigma_{\rm max}$, all three solutions are present, though again only the coolest and hottest solutions are stable (hereafter, the ``low'' and ``high'' branch, respectively). 

The existence of thermal instabilities in nature, i.e. outside the idealization of the \citet{Shakura&Sunyaev73} $\alpha$-model, remains under debate. AGN observations (e.g., \citealt{Done+07}) and some global radiation MHD simulations of radiation-dominated accretion disks do not find evidence for thermal instabilities (e.g., \citealt{Jiang+16,Jiang+19}). However, the proposed physical mechanisms responsible for quenching the instability, such as the iron opacity bump associated with high-metallicity gas (e.g., \citealt{Jiang+16}) or the presence of dynamically-strong magnetic fields (e.g., \citealt{Dexter&Begelman19,Kaur+23}) amplified and dragged in from larger scales in the disk (e.g., \citealt{Jacquemin-Ide+24}), may not exist for the mass stripped from what is likely a weakly magnetized star. While we therefore take the presence of thermal instabilities at face value in this section, the thermally-stable solutions obtained in Sec.~\ref{sec:stable} would be the relevant ones in their absence.

\begin{figure}
    \centering
    \includegraphics[width=0.49\textwidth]{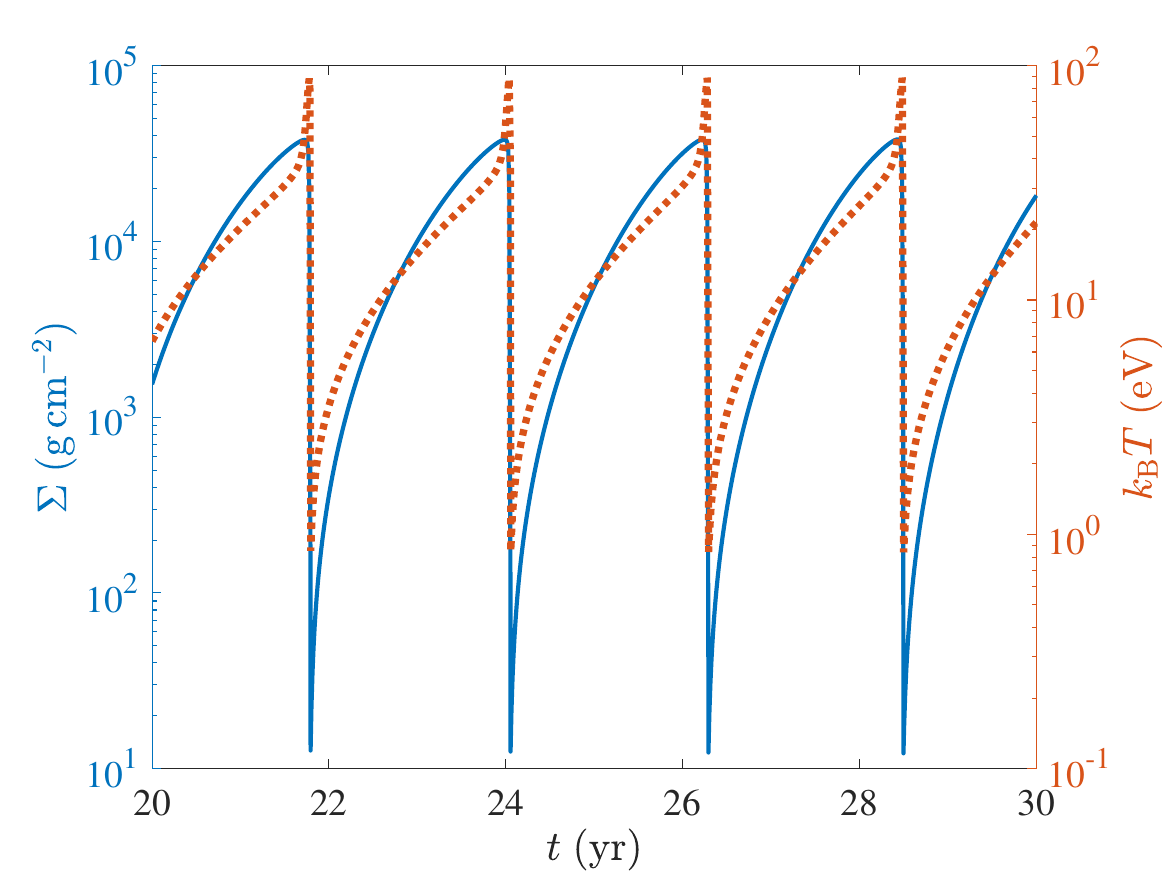}
    \includegraphics[width=0.49\textwidth]{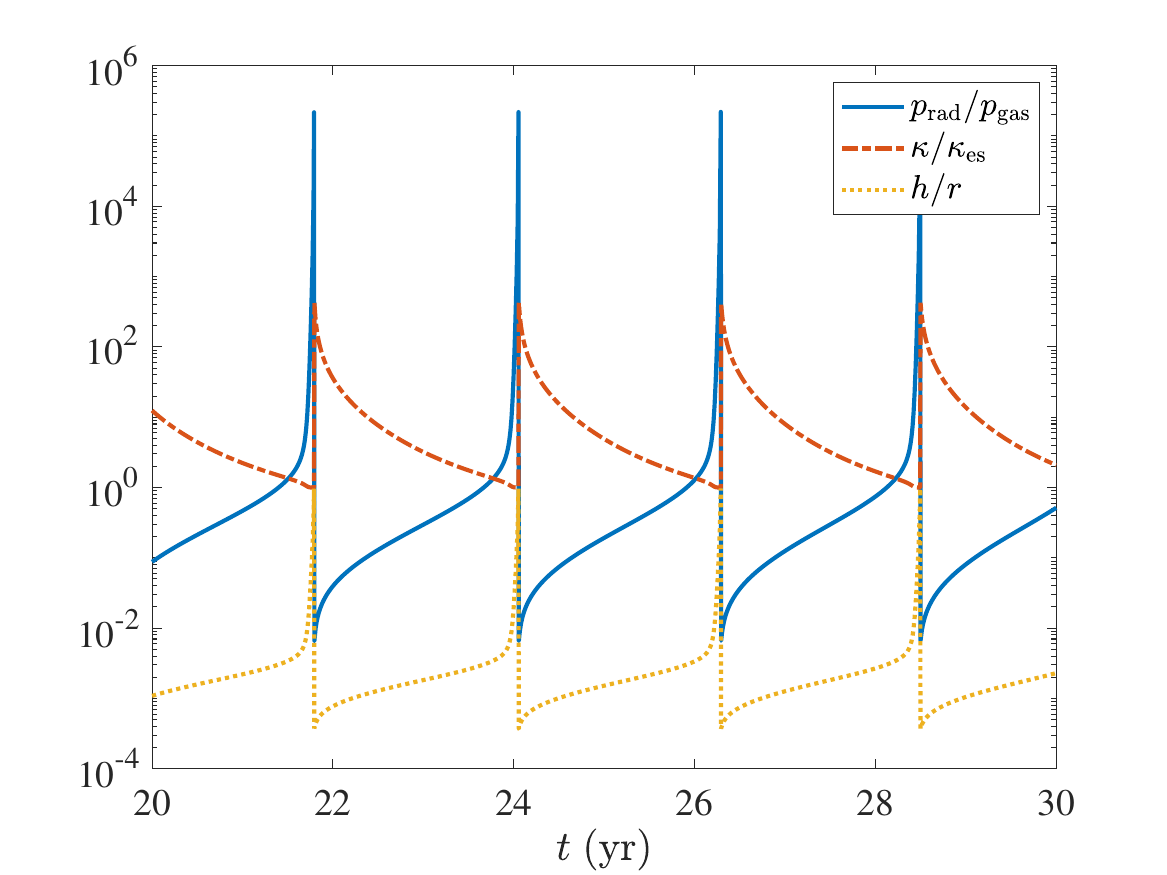}
    \includegraphics[width=0.49\textwidth]{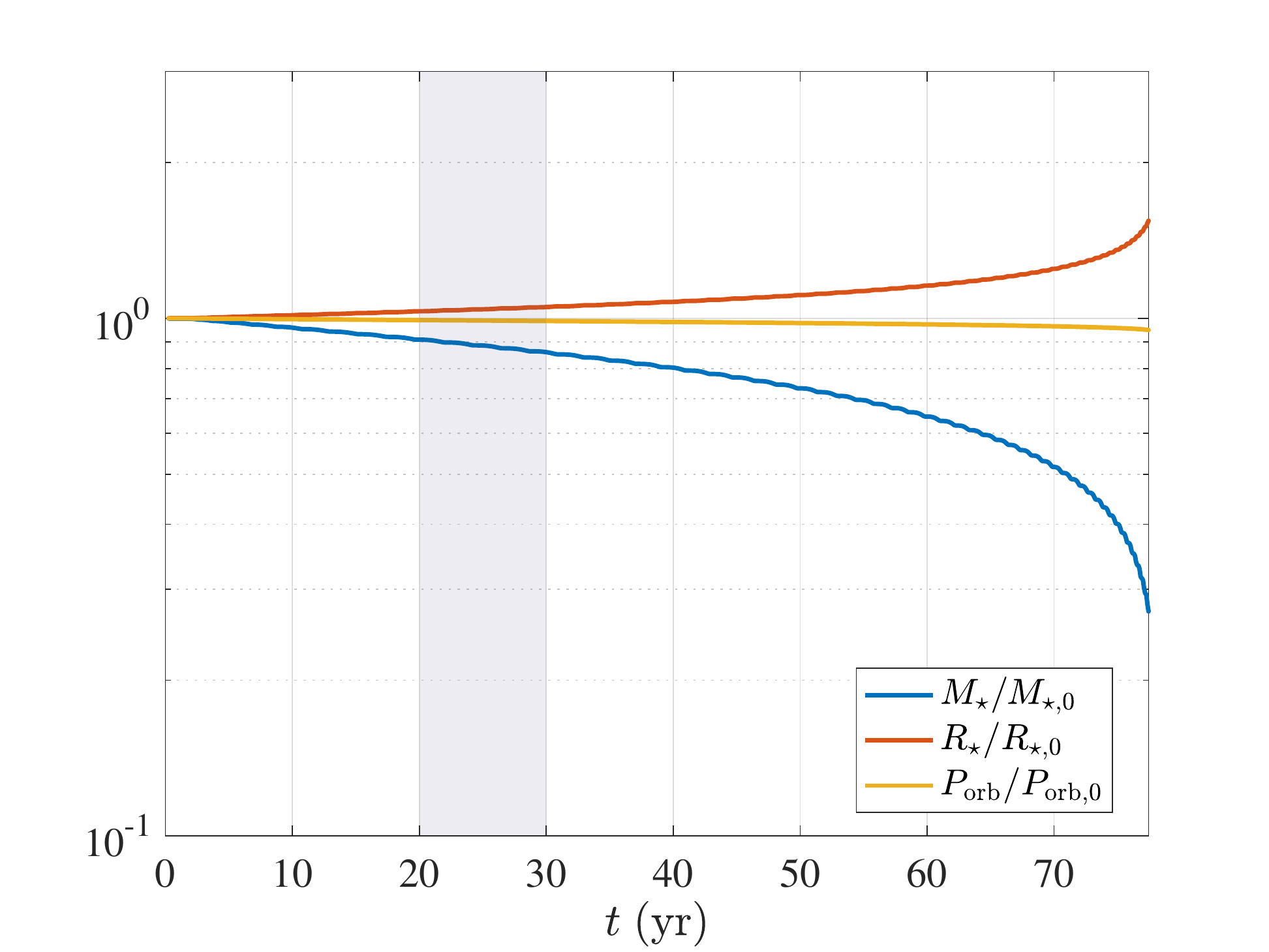}
    \includegraphics[width=0.49\textwidth]{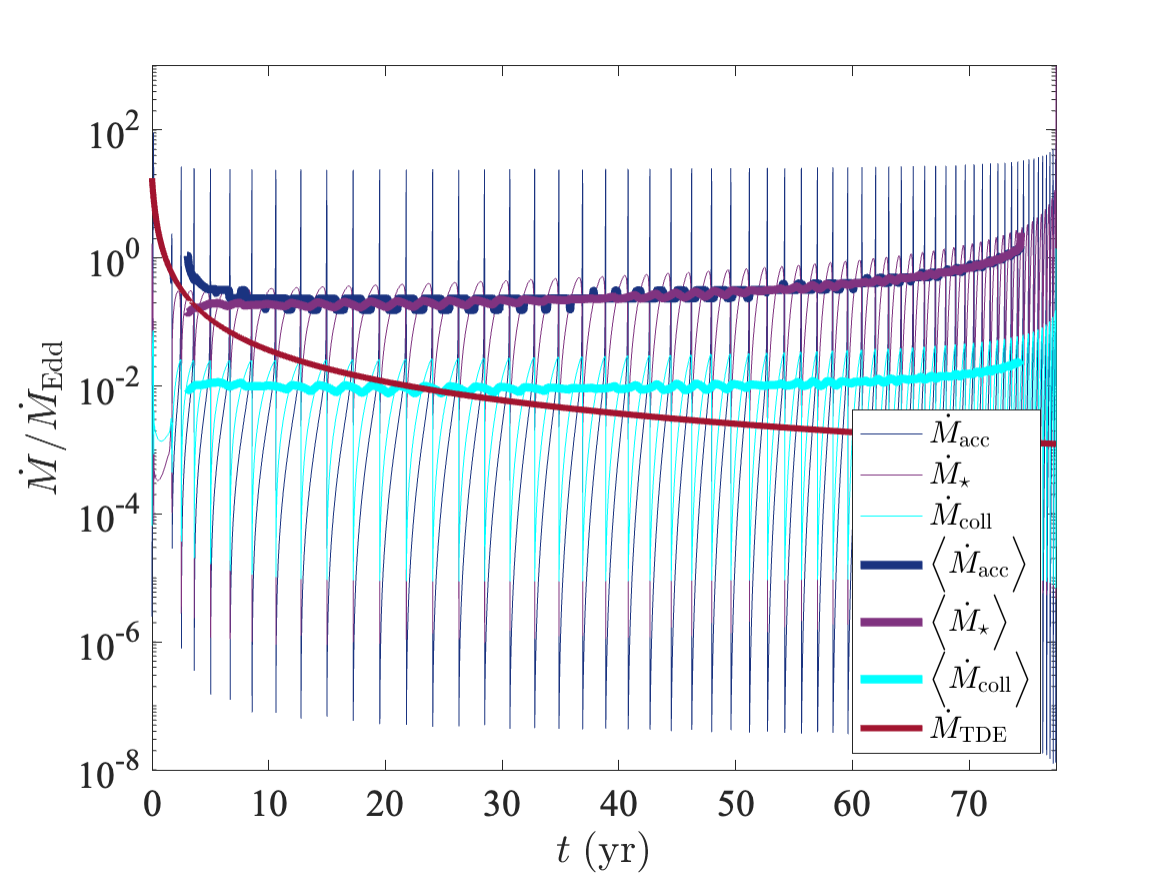}
    \caption{Disk-star evolution for an initial sun-like star ($\mathcal{M}_{\star} = \mathcal{R}_{\star} = 1$), neglecting collision/ablation heating ($\beta=0$), for fiducial parameters ($\MBH = 10^6 \, \rm M_\odot$, $P_{\rm QPE} = 4$ hr; $\eta=10^{-3}$; $\alpha = 0.1$). \textit{Top left panel}: Surface density (solid blue) and midplane temperature (dotted red) as a function of time. \textit{Top right panel:} $p_{\rm rad}/p_{\rm gas}$, $\kappa / \kappa_{\rm es}$ and $h/r$ as a function of time. Gas pressure dominates over radiation pressure and opacity is dominated by Kramer's throughout the ``low" branch of the limit cycle. \textit{Bottom left panel:} Stellar and orbital properties as a function of time ($M_\star$, $R_\star$ and $P_{\rm orb}$), from TDE to near depletion of the star. As expected for these parameters, the stellar ablation rate greatly exceeds the rate of orbital decay due to hydrodynamical drag. \textit{Bottom right panel:} Time evolution of instantaneous and cycle-averaged rates of accretion $\dot{M}_{\rm acc}$, stellar mass-stripping $|\dot{M}_{\star}|$ and disk mass impacted by the star $\dot{M}_{\rm coll}$, in comparison to the TDE fall-back rate $\dot{M}_{\rm TDE}.$ }
    \label{fig:beta0_evo}
\end{figure}
\begin{figure}
    \centering
    \includegraphics[width=0.49\textwidth]{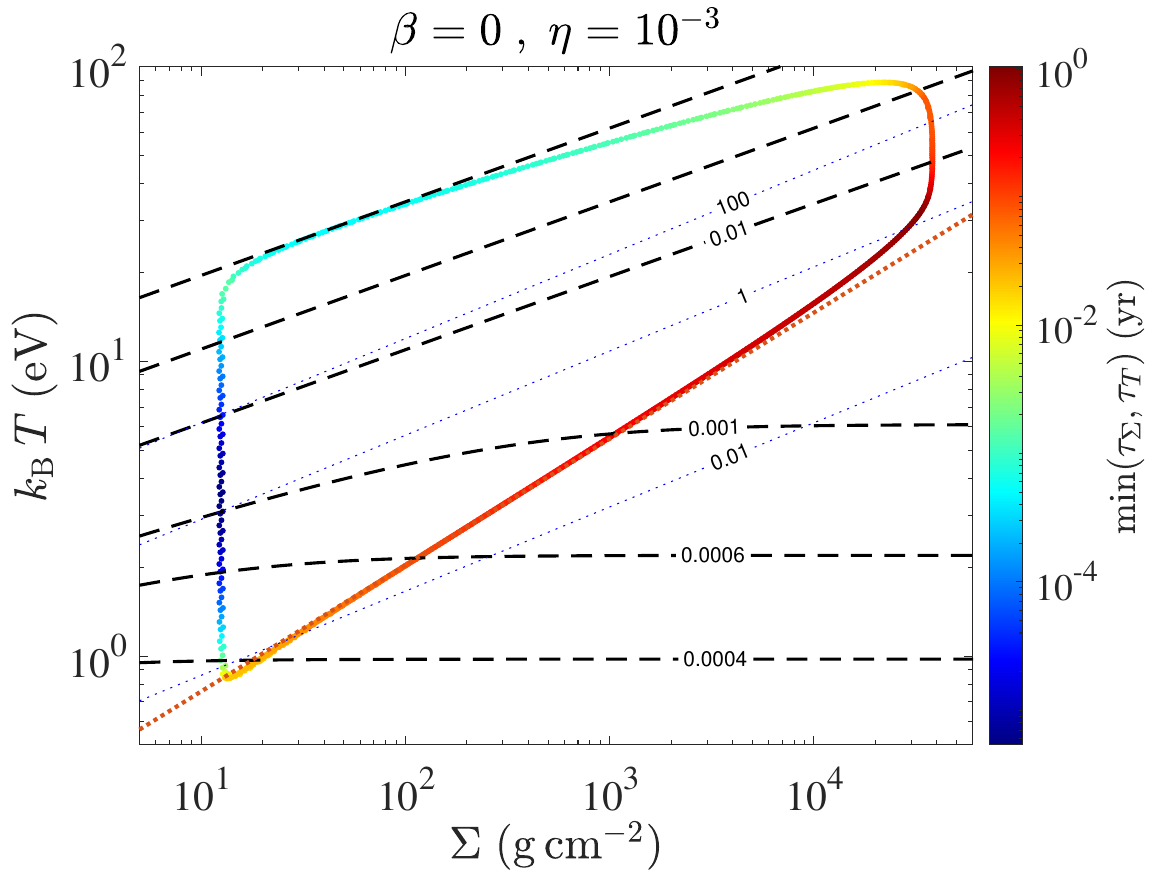}   \includegraphics[width=0.49\textwidth]{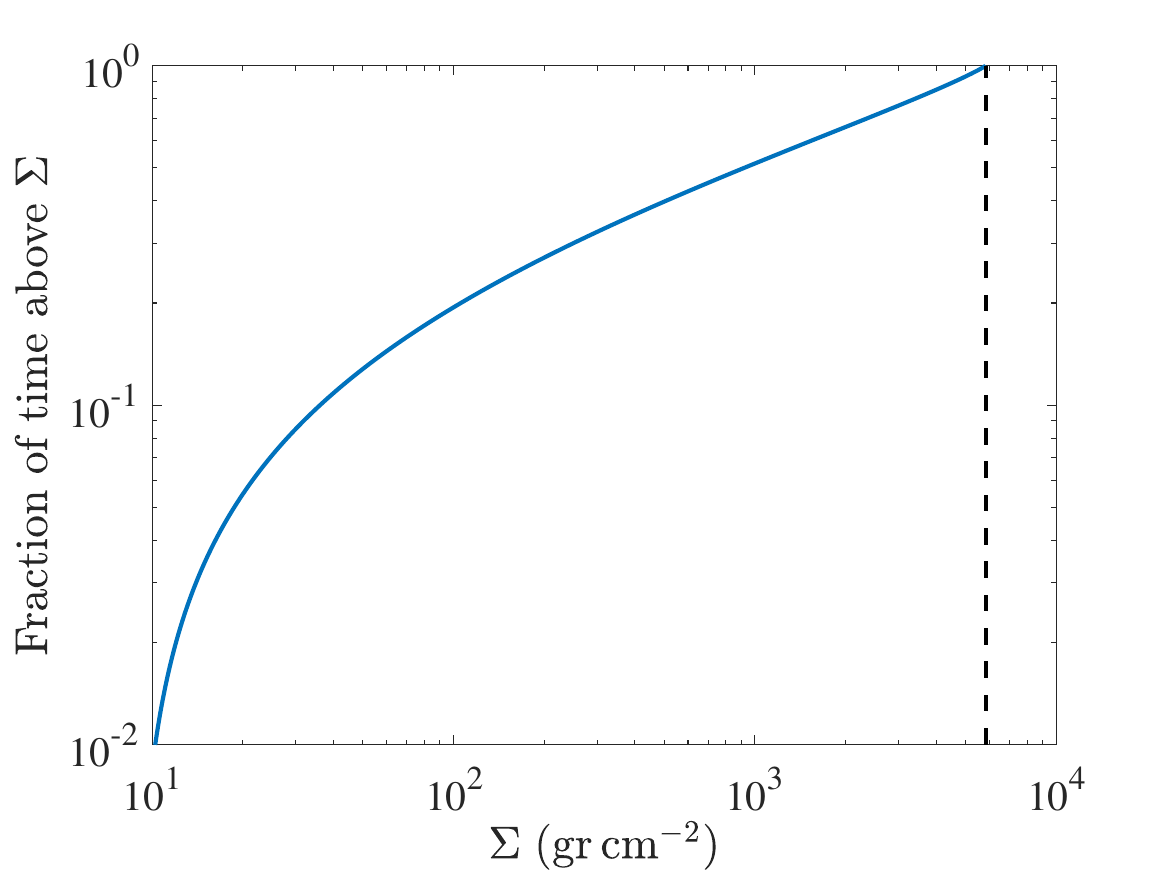}
    \includegraphics[width=0.49\textwidth]{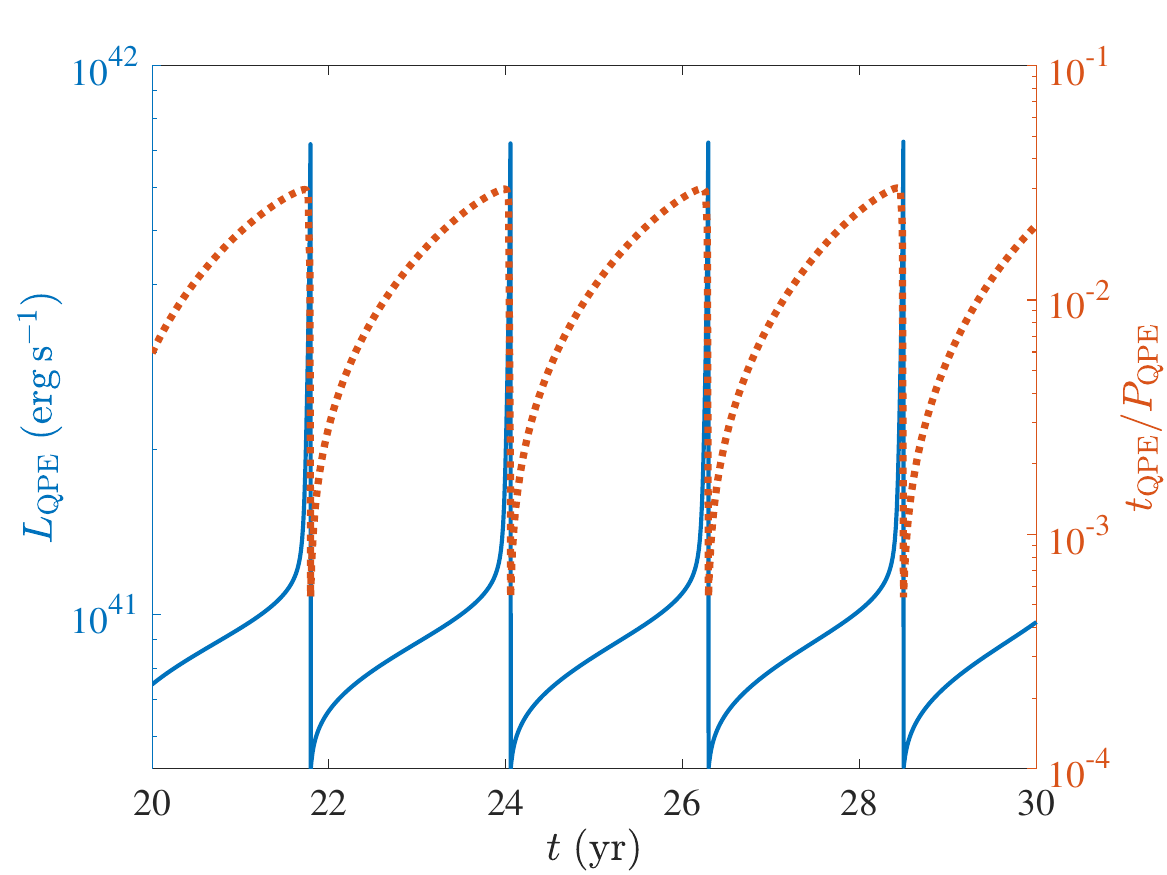}
    \caption{Disk evolution for a sun-like star orbiting a $10^{6}M_{\odot}$ SMBH, in the absence of collision/ablation heating ($\beta=0$) and $\eta=10^{-3}$. \textit{Left panel}: Limit-cycle evolution in the $T$ vs. $\Sigma$ plane. The color-map shows the system's evolution time along the limit cycle, with its maximal value obtained towards $\Sigma_{\rm max}$ along the lower thermal equilibrium branch. The \textit{black dashed} contours correspond to different values of $h/r$ (Eq.~\ref{eq:hovereq_gen}), and \textit{blue dotted} contours correspond to $p_{\rm rad}/p_{\rm gas}$. The \textit{red dotted} line is the analytical expression for the lower thermal equilibrium branch (Eq.~\ref{eq:sigma_low}). \textit{Right panel}: The fraction of time spent below $\Sigma$, as a function of $\Sigma$. \textit{Bottom panel:} QPE eruption luminosity $L_{\rm QPE}$ and duration $t_{\rm QPE}$ (expressed in ratio to the QPE period $P_{\rm QPE}$), as a function of time, following the analytic estimates of Eqs.~\eqref{eq:LQPE}, \eqref{eq:tQPE}.}
    \label{fig:beta0_limit}
\end{figure}

As illustrated by Fig.~\ref{fig:beta0_evo}, the main feature of the disk evolution for $\beta = 0$ is a limit cycle behavior in which $\Sigma$ undergoes periodic see-saw like oscillations spanning the range $\Sigma_{\rm min} < \Sigma < \Sigma_{\rm max}$.  In each cycle, the disk surface density grows gradually from low $\Sigma \approx \Sigma_{\rm min} \sim 10$ g cm$^{-2}$ until reaching $\Sigma \approx \Sigma_{\rm max} \sim 10^{4}$ g cm$^{-3}$, before rapidly dropping back to $\Sigma \approx \Sigma_{\rm min}$, starting the cycle again. The disk spends most of the time in the gas pressure-dominated low state, in which it is geometrically-thin $h/r \sim 10^{-3}$ and the viscous timescale $\tau_{\nu} \propto (h/r)^{-2}$ (Eq.~\eqref{eq:tnu}) is effectively infinite. The disk mass therefore builds up during this phase ($\dot{M}_{\rm acc} \ll |\dot{M}_{\star}| + \dot{M}_{\rm TDE}$) due to a combination of mass fall-back from the TDE (early times) and ablation from the orbiting EMRI (late times).  By contrast, once $\Sigma \approx \Sigma_{\rm max}$ the disk transitions to the high state, which is geometrically thick $h/r \sim 1$ and the viscous timescale is extremely short, causing rapid accretion ($\dot{M}_{\rm acc} \gg |\dot{M}_{\star}|+ \dot{M}_{\rm TDE})$, and rapidly reducing $\Sigma \approx \Sigma_{\rm min}.$  The duration of each cycle is shortest at early times, when the dominant source of mass being added to the disk is fall-back accretion from the TDE (see \citealt{Shen&Matzner14,XiangGruess+16}).  However, throughout most of the disk's lifetime $t \gtrsim 3$ yr, TDE fall-back accretion (or whatever process first created the gaseous disk) becomes subdominant compared to the mass added by stellar stripping, generating regular cycles every $\approx 2$ yr. This cyclic evolution continues until the star begins to lose an appreciable fraction of its total mass and starts to expand near the end of its lifetime, accelerating its destruction at around $t \approx 70$ yr.\footnote{In this particular example, the EMRI's evolution would actually terminate prior to complete ablation-induced destruction, once it overflows its Roche-lobe onto the SMBH; this occurs once $P_{\rm QPE} < P_{\rm QPE,min} \propto \mathcal{R}_{\star}^{3/2}\mathcal{M}_{\star}^{-1/2} \propto \mathcal{M}_{\star}^{-1}$ (Eq.~\eqref{eq:PQPEmin}) becomes satisfied, i.e. after only moderate fractional stellar mass-loss (since $P_{\rm QPE} \gtrsim P_{\rm QPE,min}$ in the initial state for the chosen model parameters).}

The main features of the system evolution can be understood analytically. Equating viscous heating with radiative cooling $\dot{E}_{\rm rad} \approx \dot{E}_{\rm acc}$ under the assumptions $p_{\rm rad} \ll p_{\rm gas}$ and $\kappa \simeq \kappa_{\rm K}$ gives the lower gas-pressure dominated branch (plotted as red-dotted line in the left panel of Fig.~\ref{fig:beta0_limit})
\begin{equation}
    \left. k_{\rm B} T(\Sigma) \right|_{\rm low} = k_{\rm B}\left( \frac{27}{8} \frac{A_{\kappa} \alpha}{a c} \pfrac{v_{\rm k}}{r_0}^2 \pfrac{k_{\rm B}}{\mu m_{\rm p}}^{1/2} \Sigma^3 \right)^{1/7} \simeq 15 \, {\rm eV} \;  \pfrac{\Sigma}{10^4 \, \rm g\, cm^{-2}}^{3/7} \alpha_{-1}^{1/7} \mathcal{P}_{\rm QPE,4}^{-2/7} \,,
    \label{eq:sigma_low}
\end{equation}
and a corresponding scale height of
\begin{equation}
    \left. \frac{h}{r} \right|_{\rm low} \approx \sqrt{ \frac{k_{\rm B} T}{\mu m_{\rm p} v_{\rm k}^2} } \approx 1.5\times 10^{-3} \pfrac{\Sigma}{10^4 \, \rm g\, cm^{-2}}^{3/14} \alpha_{-1}^{1/14} \,\mathcal{P}_{\rm QPE,4}^{4/21} M_{\bullet,6}^{-1/3},
    \label{eq:hoverr_low}
\end{equation}
where $\alpha_{-1} \equiv \alpha/(0.1).$
The maximal value of $\Sigma$ corresponds to where this lower branch meets the upper radiation-pressure dominated branch (for which $\kappa \simeq \kappa_{\rm T}$). Equating the analytic estimates for these two roots of $\dot{E}_{\rm rad} \approx \dot{E}_{\rm acc}$ show that they converge at a maximal surface density:
\begin{equation}
    \Sigma_{\rm max} \approx \mathcal{C}\cdot \frac{2^{33/12}}{2^{19/12}}\pfrac{r_0/v_k}{\alpha^{11}}^{1/12} \pfrac{c^{11} \mu^2 m_{\rm p}^2}{A_{\kappa}^4 k_{\rm B}^2 \kappa_{\rm es}^7 a^3}^{1/12} \approx 4\times 10^{4} \; {\rm g \, cm^{-2}} \; \alpha_{-1}^{-11/12} \mathcal{P}_{\rm QPE,4}^{1/12},
    \label{eq:sigmaadv}
\end{equation}
where we have included a prefactor $\mathcal{C} \approx 1/3$ to account for the fact that the transition between branches occurs at a somewhat lower surface density than where the two analytic branches intersect, resulting in better consistency with the maximal $\Sigma$ obtained during each cycle in our numerical calculation (Fig.~\ref{fig:beta0_evo}).

When in the low-state, accretion is negligible and the disk's surface density grows as $\dot{\Sigma} \approx |\dot{M}_{\star}|/\pi r_0^{2}$ (at late times, when $|\dot{M}_\star| \gg \dot{M}_{\rm TDE}$). The time spent at any $\Sigma_{\rm min} < \Sigma < \Sigma_{\rm max}$ can thus be estimated as
\begin{eqnarray}
\tau_{\rm \Sigma} \equiv \frac{\Sigma}{\dot{\Sigma}} = \frac{P_{\rm QPE}}{\eta}\frac{\pi R_{\star}^{2}\Sigma}{M_{\star}}\frac{p_{\star}}{p_{\rm ram}} = \frac{P_{\rm QPE}}{\eta} \pfrac{r_0}{R_\star}^4 \frac{M_{\star}}{M_{\bullet}}\left(\frac{h}{r}\right) \approx 
0.8 \,{\rm yr}\,\eta_{-3}^{-1} \pfrac{h/r}{10^{-3}} \frac{\mathcal{M}_{\star}}{\mathcal{R}_{\star}^{4}} \mathcal{P}_{\rm QPE,4}^{11/3} M_{\bullet,6}^{1/3},
\label{eq:tausigma}
\end{eqnarray}
where we have used Eqs.~\eqref{eq:dMsdt}, \eqref{eq:Mdotstar} and $\eta_{-3} \equiv \eta/(10^{-3})$. Because $\tau_{\Sigma} \propto h/r \propto \Sigma^{3/14}$ (Eq.~\eqref{eq:hoverr_low}), the disk spends roughly equal time at all $\Sigma_{\rm min} < \Sigma < \Sigma_{\rm max}$. Evaluating Eq.~\eqref{eq:tausigma} at $h/r$ (Eq.~\eqref{eq:hoverr_low}) at $\Sigma \approx \Sigma_{\rm max}$ thus gives an estimate of the total cycle duration:
\be
\tau_{\rm cyc} \approx \tau_{\Sigma_{\rm max}} \approx 1.6\,{\rm yr}\,\eta_{-3}^{-1}\alpha_{-1}^{-1/8}\frac{\mathcal{M}_{\star}}{\mathcal{R}_{\star}^{4}}\mathcal{P}_{\rm QPE,4}^{31/8}
\label{eq:taucyc}.
\ee
This roughly agrees the cycle duration $\approx 2$ yr found in our numerical calculations (Fig.~\ref{fig:beta0_evo}). 

Both the mass accretion rate $\dot{M}_{\rm acc} \propto \Sigma(h/r)^{2} \propto \Sigma^{10/7}$ peaks, and most of the total time $\tau_{\Sigma} \propto \Sigma^{3/14}$ is spent, near the highest $\Sigma \approx \Sigma_{\rm max}$ achieved during the cycle. The average accretion rate in the low-state can therefore be estimated as
\begin{eqnarray}
\langle \dot{M}_{\rm low} \rangle \simeq  3\pi \alpha r_0 v_{\rm k}\Sigma_{\rm max}\left(\frac{h}{r}\right)_{\Sigma_{\rm max}}^{2} 
\approx 4\times 10^{-3} \, \dot{M}_{\rm Edd}\alpha_{-1}^{-1/6}\frac{\mathcal{P}_{\rm QPE,4}^{5/6}}{M_{\bullet,6}}.
\end{eqnarray}
During this phase, roughly $\left< \dot{M}_{\rm low}\right> \times \tau_{\Sigma_{\rm max}} \approx 2\times 10^{-4} \, \rm M_\odot$ of mass are accreted onto the SMBH.

However, this is substantially lower than the accretion rate averaged over the many cycles, because most of the total mass is accreted during the rapid ``flushes'' that occur in the brief advective high-state phase. The latter average, more relevant to what ultimately feeds the black hole, can be estimated as the typical disk mass built up during the cycle $\sim \pi r_0^{2}\Sigma_{\rm max}$ ($\approx 10^{-2} \, M_\odot$ in this example) divided by the cycle duration: (Eq.~\eqref{eq:taucyc}), 
\begin{eqnarray}
\langle \dot{M}_{\rm acc} \rangle \approx \frac{\pi r_0^{2}\Sigma_{\rm max}}{\tau_{\rm cyc}} 
\approx 0.3 \dot{M}_{\rm Edd} \eta_{-3}\alpha_{-1}^{-19/24}\frac{\mathcal{R}_{\star}^{4}}{\mathcal{M}_{\star}}M_{\bullet,6}^{-1/3}\mathcal{P}_{\rm QPE,4}^{-59/24} \,,
\label{eq:Mdotavg}
\end{eqnarray}
again compatible with the average accretion rate from our numerical calculation (Fig.~\ref{fig:beta0_evo}).

Throughout most of the disk's evolution the accreted mass originates from the star (certainly after the initial TDE fallback rate has subsided below $\langle \dot{M}_{\rm acc} \rangle$). The timescale for the star to be destroyed can therefore be estimated as:
\be
\tau_{\rm dest} \sim \frac{3}{10}\frac{M_{\star,0}}{\langle \dot{M}_{\rm acc} \rangle} \approx \tau_{\rm cyc} \pfrac{M_\star}{\pi r_0^2 \Sigma_{\rm max}} \approx 40\,{\rm yr}\, \eta_{-3}^{-1}\alpha_{-1}^{19/24}\frac{\mathcal{M}_{\star}^{2}}{\mathcal{R}_{\star}^{4}}\frac{\mathcal{P}_{\rm QPE,4}^{59/24}}{M_{\bullet,6}^{2/3}},
\label{eq:taudest}
\ee
where the factor $\approx 3/10$ accounts for the fact that $\tau_{\rm dest} \propto \mathcal{M}_{\star}^{2}/\mathcal{R}_{\star}^{4} \propto \mathcal{M}_{\star}^{10/3}$ decreases as the star sheds mass and expands adiabatically (Eq.~\eqref{eq:Rstar}), thereby accelerating the destruction rate. Again, this agrees within a factor of $\lesssim 2$ of the stellar lifetime $\tau_{\rm dest}\approx 70$ yr found in our numerical calculation (Fig.~\ref{fig:beta0_evo}). 

\subsubsection{Moderate collision/ablation heating}
\label{sec:midbeta}

\begin{figure}
    \centering
    \includegraphics[width=0.49\textwidth]{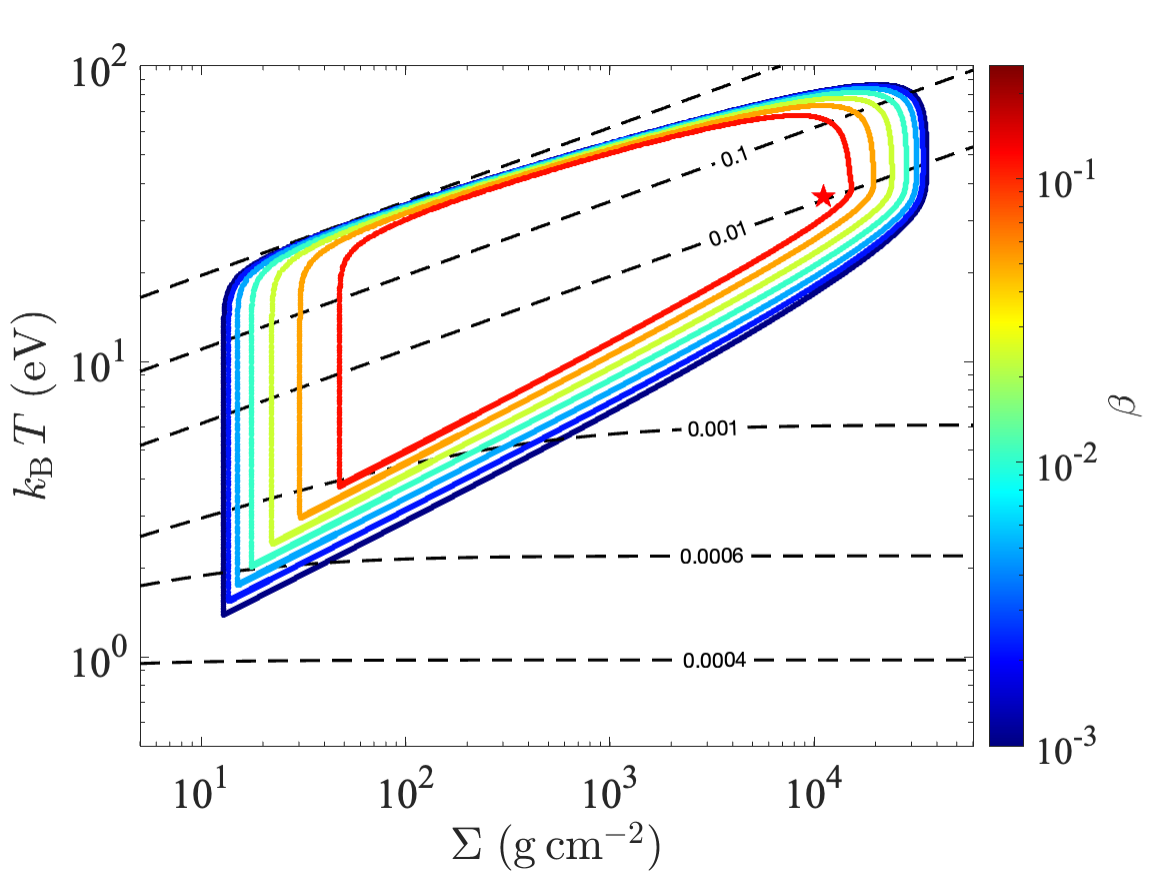}   \caption{Same as left panel of Fig.~\ref{fig:beta0_limit}, but showing evolution in the $T$ vs. $\Sigma$ plane as the value of the collision/ablation heating parameter $\beta$ is increased at constant logarithmic intervals between $10^{-3}$ to $0.25$ (as labeled in color). The thermally-stable solution obtained for $\beta = 0.25 > \beta_{\rm stable} = 3/16$ is indicated by a red star.}
    \label{fig:loops}
\end{figure}

The addition of collision/ablation heating acts to stabilize the disk evolution, even when radiation pressure dominates. This is because while the viscous heating $\dot{E}_{\rm acc} \propto (h/r)^{2} \propto p^{2}$ is a sensitive increasing function of the disk's midplane pressure $p$, the collision/ablation heating rate $\dot{E}_{\rm abl} \propto \dot{M}_{\star} \propto p_{\rm ram} \propto \Sigma/h \propto p^{-1}$ {\it decreases} with increasing midplane pressure $p$.  In Appendix \ref{sec:stability} we show that for $\beta > \beta_{\rm stable} \approx 3/16$, collision/ablation heating is sufficient to stabilize the disk in a steady-state $\dot{M}_{\rm acc} \approx |\dot{M}_{\star}|$ (we illustrate this regime explicitly in Sec.~\ref{sec:stable}). 
 
However, for intermediate efficiencies $0 \ll \beta \lesssim \beta_{\rm stable},$ collision/ablation heating is only able to stabilize the disk for a portion of its limit cycle. This is illustrated by Fig.~\ref{fig:loops}, which shows the limit cycle evolution in the $\Sigma-T$ space, for models otherwise similar to the $\beta = 0$ case presented in the previous section but now for a range of different $\beta \in [10^{-3},0.25].$  While the evolution qualitatively resembles the zero collision/ablation-heating case ($\beta = 0$; left panel of Fig.~\ref{fig:beta0_evo}), the ranges of the cycle shrink, e.g., $\Sigma_{\rm min}$ increases and $\Sigma_{\rm max}$ decreases, with increasing $\beta$. For $\beta \gtrsim \beta_{\rm stable} = 3/16,$ the cycles cease all together and the solution sets on a unique thermally-stable location (the $\beta = 0.25$ case is shown with an asterisk in Fig.~\ref{fig:loops}), as described in the next section. 

As in the $\beta = 0$ case, our numerical results for $0 \ll \beta < \beta_{\rm stable}$ can also be understood analytically. In Appendix \ref{sec:analytics_intermediate_beta}, we present estimates for $\Sigma_{\rm max}, \langle \dot{M}_{\rm acc} \rangle$, $\tau_{\rm dest}$, similar to those derived in Sec.~\ref{sec:beta0} for $\beta = 0$. The key physical difference is that the maximum surface density in the low-state $\Sigma_{\rm max}$ is no longer set by when the gas-pressure-dominated branch vanishes, but instead by the surface density/temperature above which accretion heating exceeds collision/ablation heating. In other words, while even moderate levels collision/ablation heating can dominate over accretion heating at low $\Sigma, T$ and allow the disk to evolve stably at the beginning of the cycle (despite being radiation-dominated), as $\Sigma, T$ grows, accretion heating eventually comes to dominate and the disk becomes unstable at $\Sigma_{\rm max}$.

\subsection{Thermally Stable Evolution}

\label{sec:stable}
\begin{figure}
    \centering
    \includegraphics[width=0.49\textwidth]{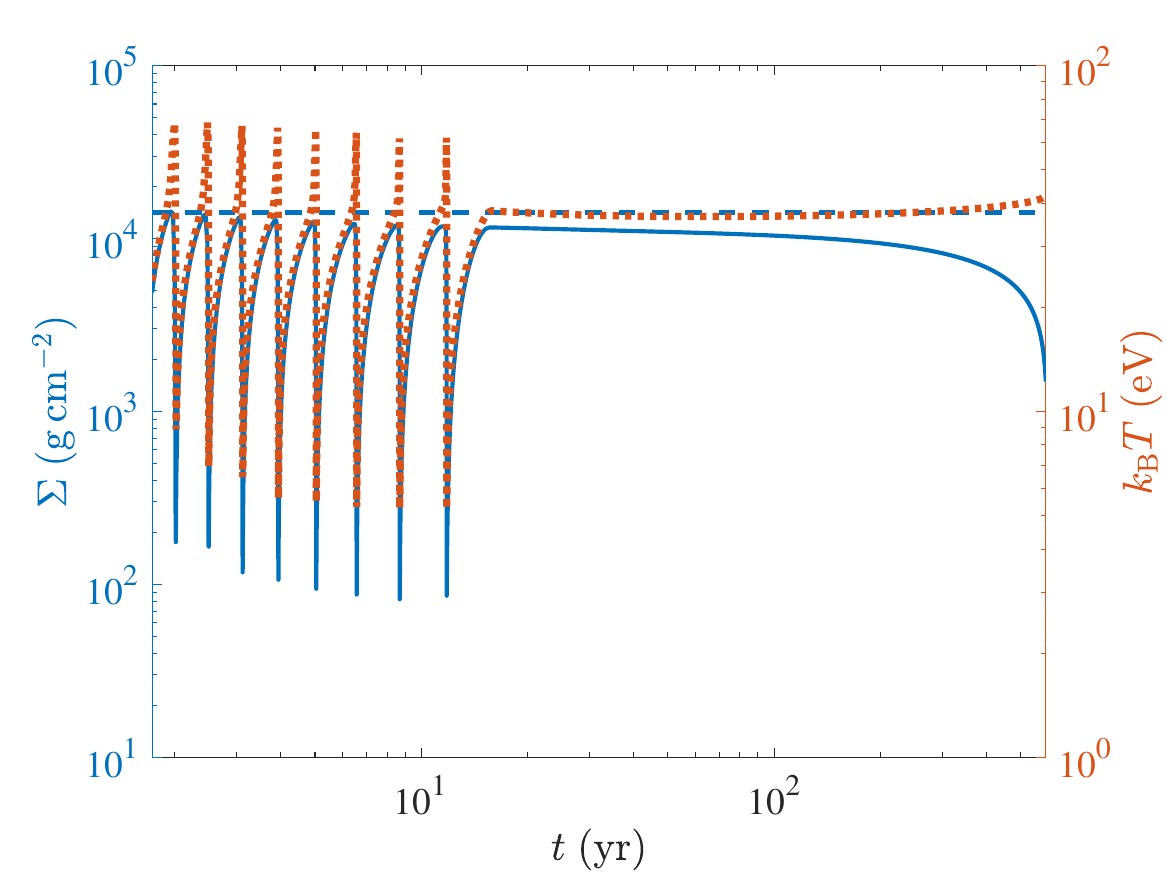}
    \includegraphics[width=0.49\textwidth]{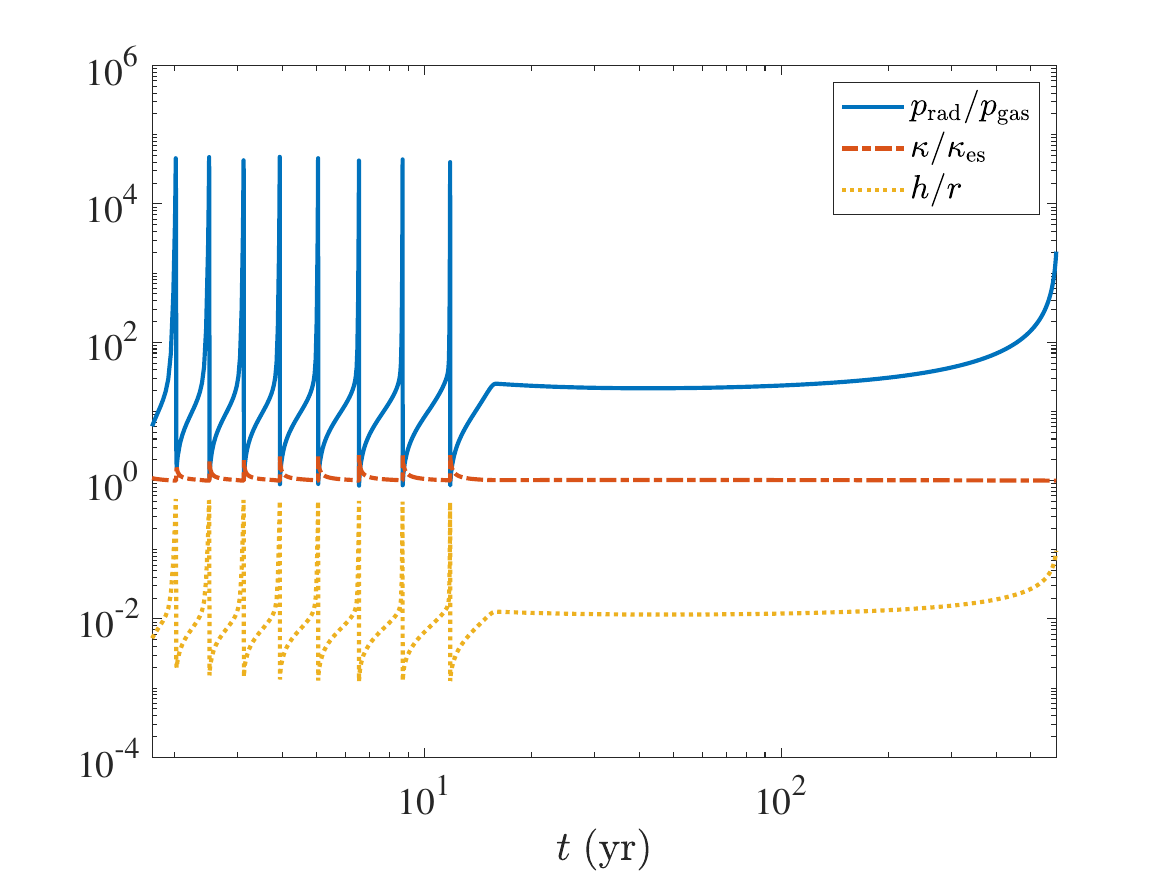}
    \includegraphics[width=0.49\textwidth]{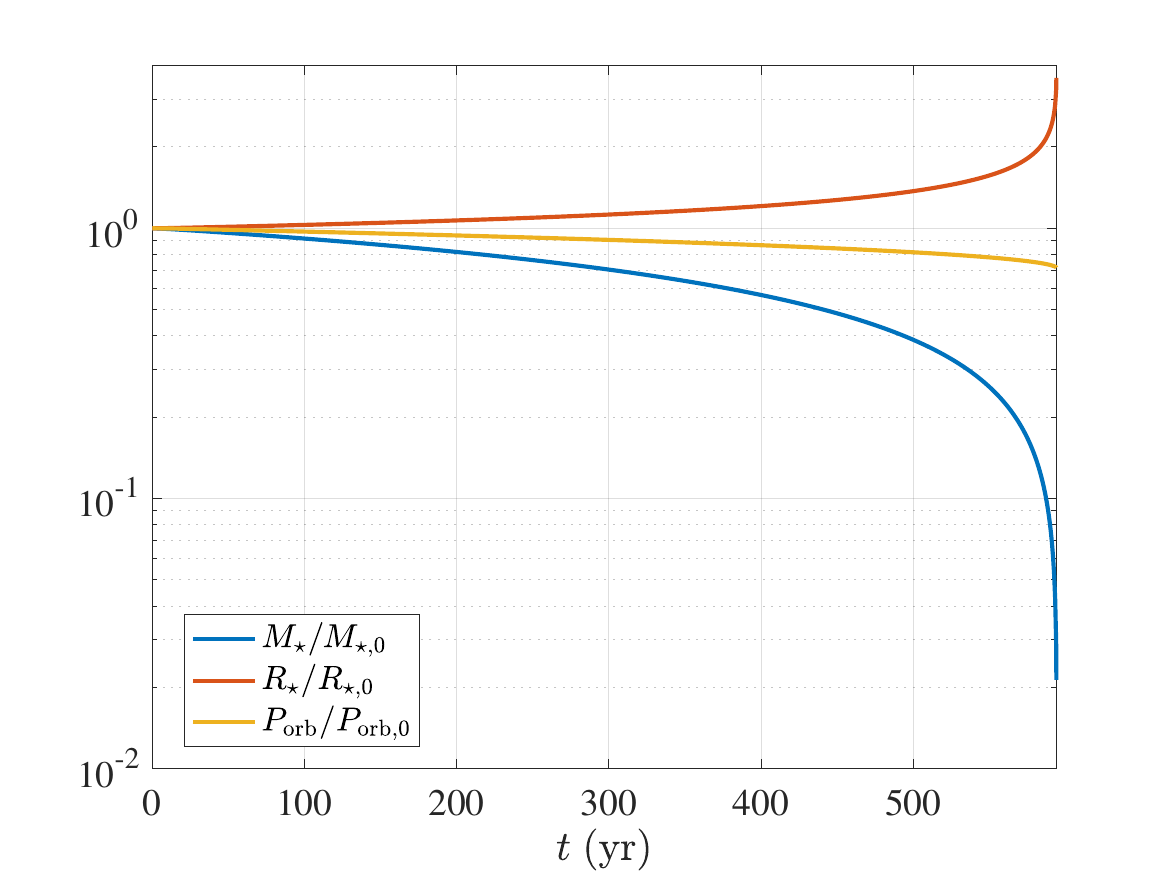}
    \includegraphics[width=0.49\textwidth]{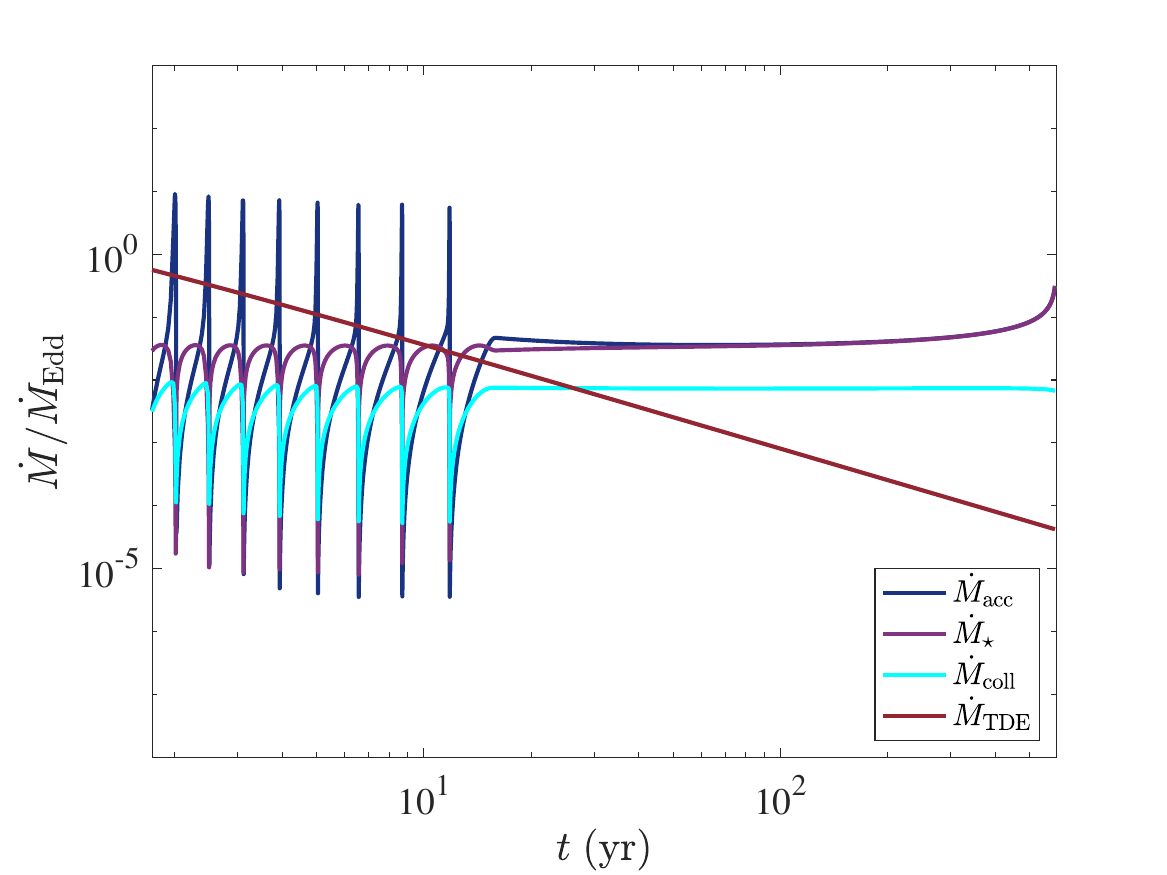}
    \caption{Same as Fig.~\ref{fig:beta0_evo}, but in the presence of substantial collisions/ablation heating, with $\beta=0.25$. \textit{Top left panel}: Surface density (solid blue) and midplane temperature (dotted red) initially fluctuate when accretion heating due to the TDE fallback is the dominant energy source, but settle onto a thermally stable evolution throughout most of the system's lifetime after collisional/ablation heating from the EMRI comes to dominate. Dashed blue horizontal line is the analytical estimate of the equilibrium $\Sigma_{\rm TS}$ (Eq.~\eqref{eq:sigmastable}), evaluated for the initial stellar and orbital properties. \textit{Top right panel}: As opposed to the low branch of the $\beta=0$ regime, here $p_{\rm rad}\gg p_{\rm gas}$ and $\kappa\approx \kappa_{\rm es}$ during the thermally stable phase of the system's evolution. \textit{Bottom left panel}: Stellar and orbital properties as a function of time. The star is completely ablated after $\sim 600 \, \rm yr$, in agreement with the analytical estimate of Eq.~\eqref{eq:taudestST}. \textit{Bottom right panel}: Instantaneous rates of accretion $\dot{M}_{\rm acc}$, stellar ablation $|\dot{M}_\star|$ and disk mass impacted by the star $\dot{M}_{\rm coll}$, in comparison to the TDE fallback rate $\dot{M}_{\rm TDE}$. The transition to a thermally stable evolution occurs roughly when the TDE fallback rate drops below the equilibrium accretion rate (Eq.~\ref{eq:MdotTS}).}
    \label{fig:beta1}
\end{figure}

For sufficiently high $\beta \gtrsim \beta_{\rm stable} = 3/16,$ collision/ablation heating is sufficient to stabilize the disk's thermal evolution. Our numerical calculation with $\beta = 1$, shown in Fig.~\ref{fig:beta1}, confirm this behavior. Although the disk still undergoes limit-cycle behavior at early times when $\dot{M}_{\rm TDE} \approx \dot{M}_{\rm acc} \gg |\dot{M}_{\star}|,$ once $\dot{M}_{\rm acc} \approx |\dot{M}_{\star}|$ is achieved after $t \gtrsim {20}$ yr, the accretion rate and surface density settle at roughly constant values $\dot{M}_{\rm acc} \approx {\ 0.04\dot{M}_{\rm Edd}}$ and $\Sigma \approx 10^{4}$ g cm$^{-2}$, respectively.  Both are intermediate to those obtained in the unstable limit cycle evolution (analogous to the solution shown with a star in Fig.~\ref{fig:loops}). Radiation pressure dominates over gas pressure in this state {\bf ($p_{\rm rad}/p_{\rm gas} \gtrsim 20$)} and electron scattering dominates the opacity, but unlike in the advectively-cooled high state, the disk is radiatively-cooled and geometrically thin $ h/r \sim {10^{-2}}$. 

The disk properties in this ``collision-supported'' state can be derived analytically. Unlike in the limit-cycle evolution, the accretion rate is approximately constant once a steady state is reached, the disk scale-height can be simply estimated by equating the accretion rate $\dot{M}_{\rm acc}$ (Eq.~\eqref{eq:Mdotacc}) with the mass-stripping rate $|\dot{M}_{\star}|$ (Eq.~\eqref{eq:Mdotstar}). Because both scale linearly with the disk surface density, this gives an expression for the disk scale-height:
\be
\frac{h}{r} = \left(\frac{2\eta \xi_{\rm v}^{2}}{3\pi \alpha}\frac{R_{\star}^{4}}{r_0^{4}}\frac{M_{\bullet}}{M_{\star}}\right)^{1/3} \approx 1.1\times 10^{-2}\left(\frac{\eta_{-3}\xi_{\rm v}^{2}}{\alpha_{-1}}\right)^{1/3}\frac{\mathcal{R}_{\star}^{4/3}}{\mathcal{M}_{\star}^{1/3}}\mathcal{P}_{\rm QPE,4}^{-8/9}M_{\bullet,6}^{-1/9} \,.
\label{eq:hoverr_stable}
\ee
Likewise, thermal equilibrium ($dE_{\rm th}/dt = 0;$ Eq.~\eqref{eq:dEthdt}) for a radiation-supported disk can be written:
\be
\left(\frac{3}{8} + \beta \right)\dot{M}_{\rm acc}v_{\rm k}^{2} = \frac{L_{\rm Edd}}{4}\left(\frac{h}{r}\right) = \frac{1}{4}\epsilon \dot{M}_{\rm Edd}c^{2}\left(\frac{h}{r}\right),
\ee
giving a second equation for the disk scale-height:
\be
\frac{h}{r} \simeq \frac{3}{2}\frac{(1+ \frac{8}{3}\beta)}{\epsilon}\frac{R_{\rm g}}{r_0}\frac{\dot{M}_{\rm acc}}{\dot{M}_{\rm Edd}},
\label{eq:hoverrB}
\ee
where $\epsilon \equiv L_{\rm Edd}/\dot{M}_{\rm Edd}c^{2} = 0.1.$ Equating \eqref{eq:hoverr_stable} with \eqref{eq:hoverrB} gives the accretion rate:
\be
\frac{\dot{M}_{\rm acc}}{\dot{M}_{\rm Edd}} = \frac{2}{3}\frac{\epsilon}{(1 + \frac{8}{3}\beta)}\frac{r_0}{R_{\rm g}}\frac{h}{r} = \frac{0.07}{1 + \frac{8}{3}\beta }\left(\frac{\eta_{-3}\xi_{\rm v}^{2}}{\alpha_{-1}}\right)^{1/3}\frac{\mathcal{R}_{\star}^{4/3}}{\mathcal{M}_{\star}^{1/3}}\mathcal{P}_{\rm QPE,4}^{-2/9}M_{\bullet,6}^{-7/9}.
\label{eq:MdotTS}
\ee
This equilibrium accretion rate is established on the viscous timescale (Eq.~\eqref{eq:tnu}),
\begin{eqnarray}
\tau_{\nu} \simeq \frac{1}{3\pi}\frac{P_{\rm QPE}}{\alpha}\left(\frac{h}{r}\right)^{-2} \approx 4.4\,{\rm yr}\,\, \alpha_{-1}^{-1/3}\eta_{-3}^{-2/3}\xi_{\rm v}^{-4/3}\frac{\mathcal{M}_{\star}^{2/3}}{\mathcal{R}_{\star}^{8/3}}\mathcal{P}_{\rm QPE,4}^{25/9}M_{\bullet,6}^{2/9} \,.
\label{eq:tvisc}
\end{eqnarray}
The steady-state disk surface density is given by 
\begin{eqnarray}
\Sigma_{\rm TS} &=& \frac{\dot{M}_{\rm acc}}{3\pi \alpha r_0 v_{\rm k} (h/r)^{2}} 
\approx 2.4\times 10^{4}\,{\rm g\,cm^{-2}}\, \frac{1}{1 + \frac{8}{3}\beta }(\eta_{-3}\xi_{\rm v}^{2})^{-1/3}\alpha_{-1}^{-2/3}\frac{\mathcal{M}_{\star}^{1/3}}{\mathcal{R}_{\star}^{4/3}}\frac{\mathcal{P}_{\rm QPE,4}^{11/9}}{M_{\bullet,6}^{2/9}}.
\label{eq:sigmastable}
\end{eqnarray}
As a consistency check, this is seen to match Eq.~\eqref{eq:sigma_intermediate} for $\beta = \beta_{\rm stable} = 3/16$.

Accreting at $\dot{M}_{\rm acc}$ (Eq.~\eqref{eq:MdotTS}), the star destruction time is approximately given by
\be
\tau_{\rm dest} \approx \frac{9}{16}\frac{M_{\star,0}}{\dot{M}_{\rm acc}} \approx 320\,{\rm yr}\, \left(1 + \frac{8}{3}\beta \right)\left(\frac{\alpha_{-1}}{\eta_{-3}\xi_{\rm v}^{2}}\right)^{1/3}\frac{\mathcal{M}_{\star}^{4/3}}{\mathcal{R}_{\star}^{4/3}}\mathcal{P}_{\rm QPE,4}^{2/9}M_{\bullet,6}^{-2/9} \,.
\label{eq:taudestST}
\ee
Again, the factor $\approx 9/16$ accounts for the fact that $\tau_{\rm dest} \propto \mathcal{M}_\star^{16/9}$ decreases as the star loses mass.

\subsection{Thermally Stable Evolution at Long Periods}
\label{sec:longperiod}

\begin{figure}
    \centering
    \includegraphics[width=0.49\textwidth]{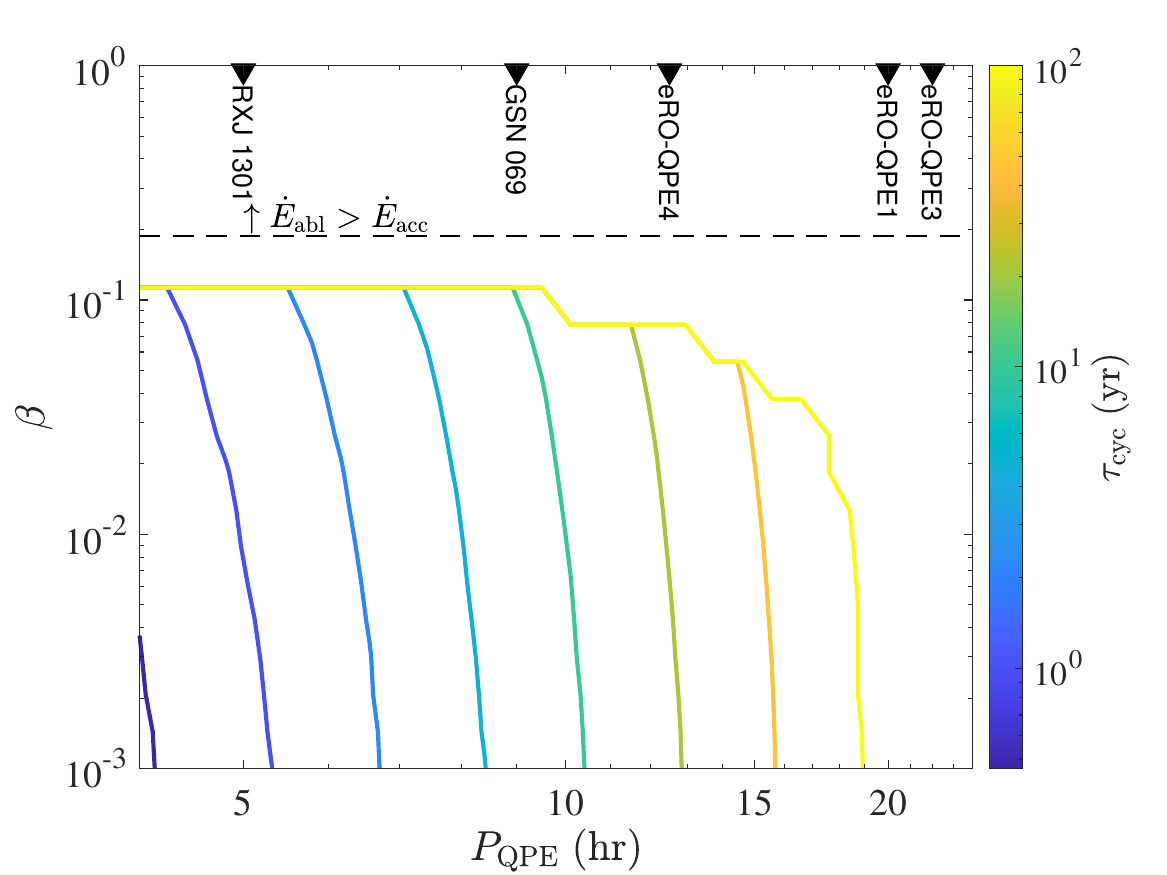}
    \includegraphics[width=0.49\textwidth]{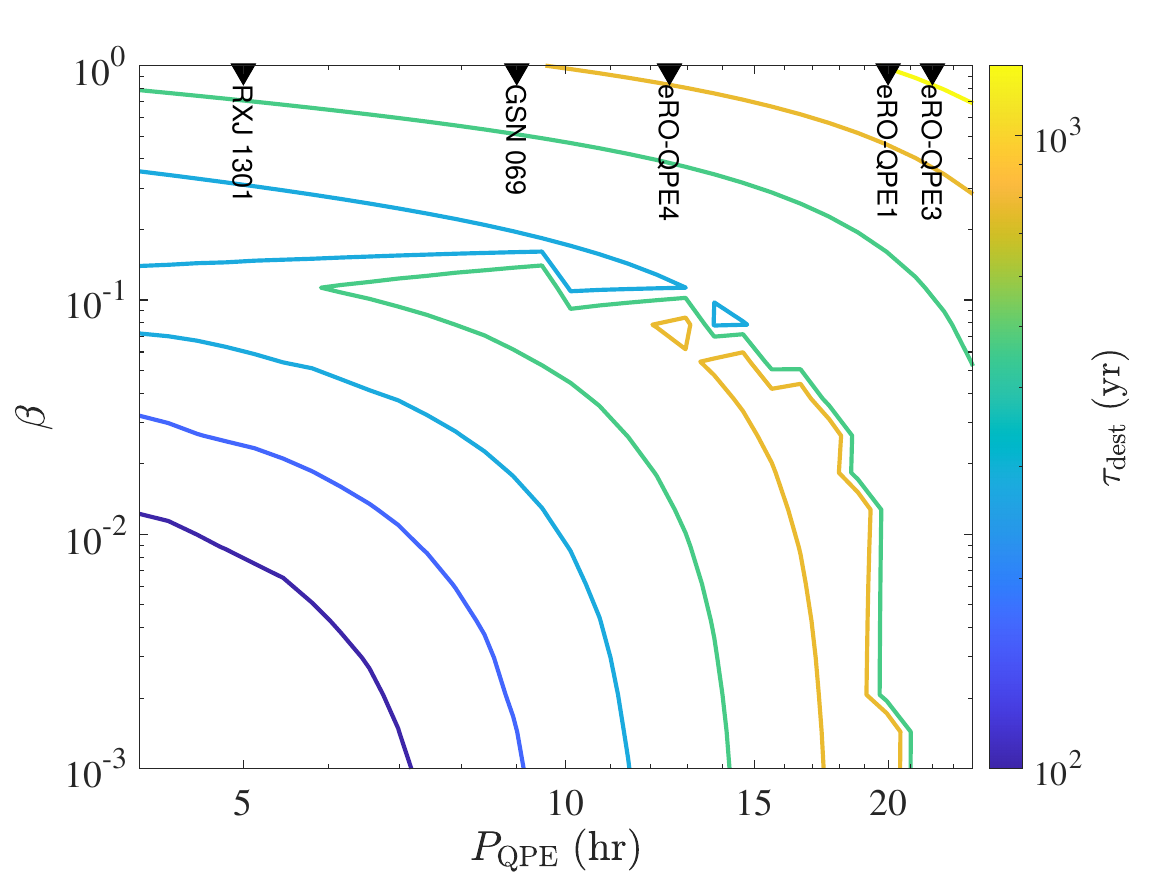}
    \includegraphics[width=0.49\textwidth]{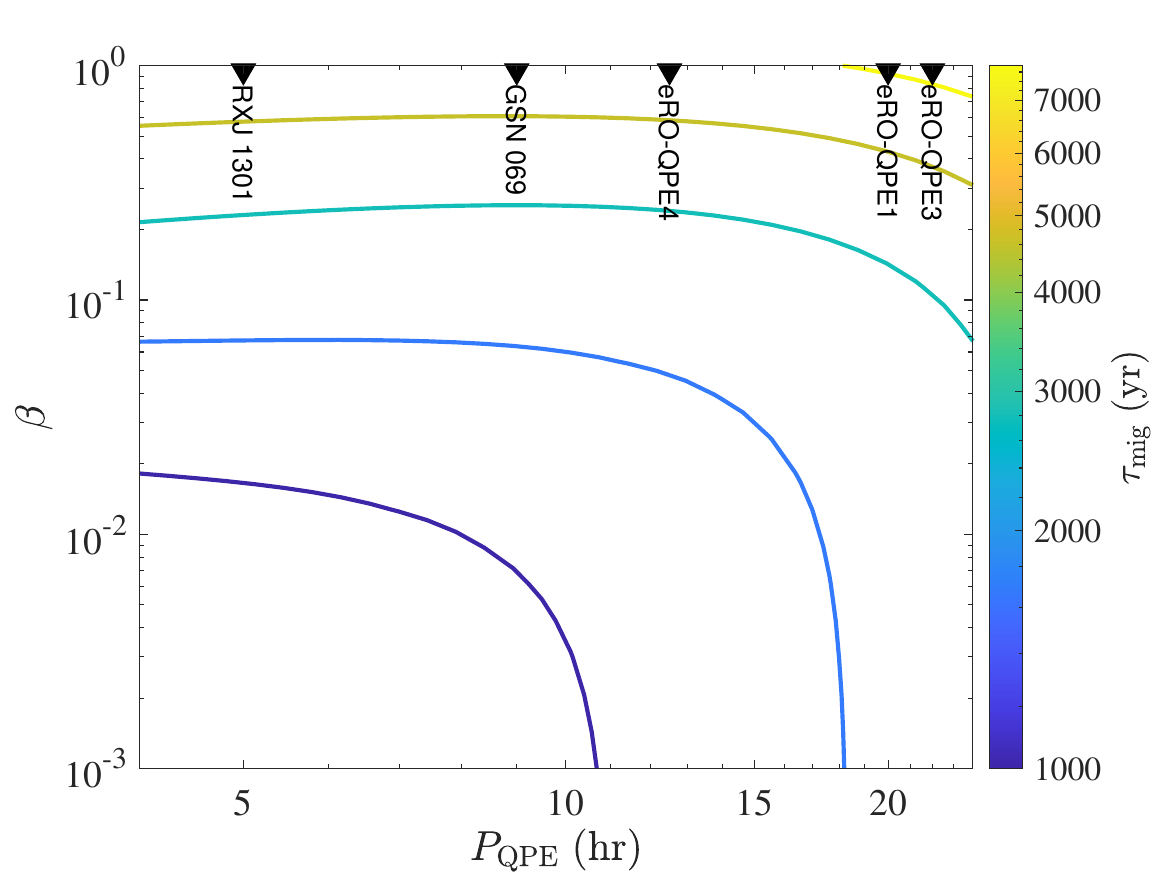}
    \includegraphics[width=0.49\textwidth]{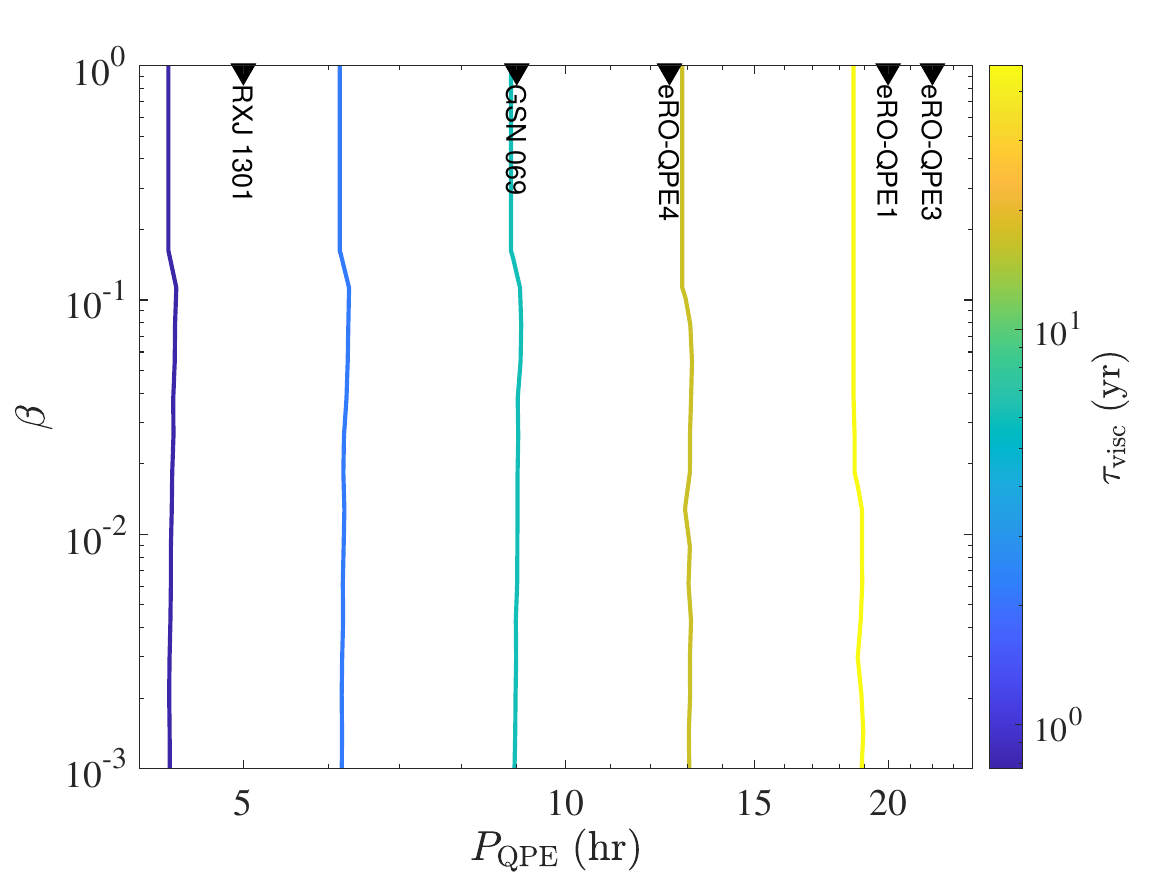}
    \caption{Contours show various key timescales as a function of (half) the orbital period $P_{\rm QPE}$ and the collisional/ablation heating efficiency $\beta$. The periods of some known QPE sources are shown along the top axis. \textit{Top left panel}: Cycle duration (in the case of a thermally unstable solution). For $\beta>\beta_{\rm stable}$ or $p_{\rm gas}>p_{\rm rad}$, the system evolves to a thermal equilibrium, and limit cycles disappear (the boundary denoted by the yellow line). \textit{Top right panel}: Stellar destruction time due to mass-stripping from disk collisions. \textit{Bottom left panel}: Migration time due to disk-induced drag. \textit{Bottom right panel}: Disk viscous time.}
    \label{fig:beta_P_QPE_contours}
\end{figure}

Thermally stable evolution may be achieved even in the absence of collision/ablation heating ($\beta \approx 0$), for sufficiently long orbital periods. At these wider radii, $p_{\rm rad} \lesssim p_{\rm gas}$, and the system evolves to a radiatively cooling equilibrium, being fed by the mass collisionally stripped from the star.  
Fig.~\ref{fig:beta_P_QPE_contours} summarizes the key timescales of the star-disk system as a function of $\beta$ and $\mathcal{P}_{\rm QPE}$, for $R_{\star} = 2 R_{\odot}$ and otherwise fiducial parameters. For $\beta\lesssim 3/16$, we see that thermal stability is obtained for $P_{\rm orb} \simeq 2P_{\rm QPE} \gtrsim 18 \, \rm hr$. 

\section{Implications}
\label{sec:discussion}

\subsection{Secular Evolution of the EMRI Orbit}
\label{sec:orbit}

Our model neglects any changes to the stellar orbit, due to hydrodynamical or gravitational interactions with the disk, implicitly assuming that they occur slower than the star's destruction. We now explore when this is a good approximation. 

The timescale required for the stellar orbit to undergo appreciable radial migration, or to be ground down into the disk midplane, via gas drag is approximately given by that required to intercept its own mass in disk material, $\tau_{\rm mig}$ (e.g., \citealt{MacLeod&Lin20,Generozov&Perets23,Linial&Quataert24}). 
In cases when the disk exhibits limit-cycle evolution and spends most of its evolution near the surface density $\Sigma_{\rm max}$ (Sec.~\ref{sec:beta0}, \ref{sec:midbeta}), the star sweeps up (or rather, collides with) mass at the rate 
\be
\dot{M}_{\rm coll} \approx \frac{\pi R_{\star}^{2}\Sigma_{\rm max}f_{\Sigma_{\rm max}}}{P_{\rm QPE}},
\ee
where the factor $f_{\rm \Sigma_{\rm max}} \sim {\rm ln}\left(\Sigma_{\rm max}/\Sigma_{\rm min}\right)^{-1} \approx 0.1$ accounts for the fraction of the limit-cycle spent at $\Sigma \approx \Sigma_{\rm max}$.  The migration timescale is thus given by:
\begin{eqnarray}
\tau_{\rm mig} \approx \frac{M_{\star}}{\dot{M}_{\rm coll}} \approx \frac{r_0^{2}}{R_{\star}^{2}}\frac{\tau_{\rm dest}}{\tau_{\rm cyc}}\frac{P_{\rm QPE}}{f_{\Sigma_{\rm max}}},
\label{eq:tausw}
\end{eqnarray}
and hence relative to the destruction time,
\be
\frac{\tau_{\rm mig}}{\tau_{\rm dest}} \approx \frac{r_0^{2}}{R_{\star}^{2}}\frac{P_{\rm QPE}}{f_{\Sigma_{\rm max}}\tau_{\rm cyc}} \approx 120 \eta_{-3}\alpha_{-1}^{1/8}\frac{\mathcal{R}_{\star}^{2}}{\mathcal{M}_{\star}}\frac{M_{\bullet,6}^{2/3}}{\mathcal{P}_{\rm QPE,4}^{37/24}} ,
\label{eq:taumig}
\ee
where we have made use of the fact that $\tau_{\rm dest} \approx M_{\star}/\langle \dot{M}_{\rm acc} \rangle$ and $\langle \dot{M}_{\rm acc} \rangle \approx \pi r_0^{2}\Sigma_{\rm max}/\tau_{\rm cyc}$ and in the final equality have used Eq.~\eqref{eq:taucyc} for $\tau_{\rm cyc}$ in the $\beta = 0$ limit (qualitatively similar results are obtained using Eq.~\eqref{eq:taucyc2} for $\tau_{\rm cyc}$). The above ratio only grows as the star loses mass, as $R_\star^2/M_\star \propto M_\star^{-5/3}$ for adiabatic expansion $R_{\star} \propto M_{\star}^{-1/3}$ (Eq.~\eqref{eq:Rstar}).

In the thermally-stable case ($\beta > 3/16$; Sec.~\ref{sec:stable}), the surface density of the disk is approximately constant in time $\Sigma \approx \Sigma_{\rm TS}$ (Eq.~\eqref{eq:sigmastable}) and the migration to destruction timescale ratio be written (now taking $f_{\Sigma_{\rm max}} = 1$)
\be
\frac{\tau_{\rm mig}}{\tau_{\rm dest}} \approx \frac{M_{\star}P_{\rm QPE}}{\pi R_{\star}^{2}\Sigma_{\rm TS}\tau_{\rm dest}} \approx 9 (\eta_{-3}\xi_{\rm v}^{2})^{2/3} \alpha_{-1}^{1/3} \frac{\mathcal{R}_{\star}^{2/3}}{\mathcal{M}_{\star}^{2/3}}\frac{M_{\bullet,6}^{4/9}}{\mathcal{P}_{\rm QPE,4}^{4/9}},
\label{eq:taumig_ST}
\ee
where we have used Eq.~\eqref{eq:taudestST} for $\tau_{\rm dest}$. As in the thermally unstable regime, the timescale ratio $\tau_{\rm mig}/\tau_{\rm dest}$ increases as the star loses mass and expands, as the ratio scales as $(R_\star/M_\star)^{2/3} \propto M_\star^{-8/9}$.

Thus, solar-type stars $\mathcal{M}_{\star} \sim \mathcal{R}_{\star} \sim 1$ on QPE-like orbital periods likely have $\tau_{\rm mig} > \tau_{\rm dest}$ for $\eta \gtrsim 10^{-4}-10^{-3}$, indicating the star will be destroyed prior to significant evolution in its orbit and justifying our model's neglect of orbital evolution. On the other hand, for smaller $\eta$ or larger orbital periods ${P}_{\rm orb} \gtrsim $ days, the star may be ground into the disk prior to being destroyed. Once the star's angular momentum aligns with that of the disk, strong collisions between the two will cease and the gaseous disk will quickly be depleted of mass. 

Much more compact stars than the sun, such as M-dwarfs ($\mathcal{R}_{\star} \approx 0.1$) or white dwarfs ($\mathcal{R}_{\star} \approx 0.01$), can also have $\tau_{\rm mig} \gg \tau_{\rm dest}$. However, in an absolute sense their destruction times $\tau_{\rm dest} \propto \mathcal{R}_{\star}^{-4}$ (Eq.~\eqref{eq:taudest}) can be orders of magnitude longer than even the gravitational-wave inspiral time, and the gaseous disks and associated QPE-like eruptions they create would be far less conspicuous.  


In addition to hydrodynamical drag, the star and the disk will exert gravitational torques on each other, potentially leading to precession and (with dissipation) eventual alignment of their orbital planes. These torques will act on a characteristic timescale
\begin{equation}
    \tau_{\rm grav} \approx \frac{\sqrt{G\MBH r_0}}{G M_\star / r_0} \sim P_{\rm QPE} \pfrac{\MBH}{M_\star} \approx 500 \, {\rm yr} \;  \frac{M_{\bullet,6} \mathcal{P}_{\rm QPE,4}}{\mathcal{M}_\star} \,, 
    \label{eq:taugrav}
\end{equation}
where the mass of the disk has been safely neglected relative to that of the star. Though comparable to the stellar destruction time $\tau_{\rm dest}$, the timescale for gravitational torques is much slower than other processes shaping the stellar/disk system such as Lense-Thirring precession (Eq.~\eqref{eq:Tprec}), and so we do not expect gravitational torques to significantly alter the system evolution.

As summarized in Table \ref{tab:timescales} and Fig.~\ref{fig:beta_P_QPE_contours}, for fiducial parameters and solar-type stars, we find a typical hierarchy of timescales: \be
P_{\rm QPE} \ll \tau_{\rm prec} \ll \tau_{\rm cyc} \ll \tau_{\rm dest} \ll \tau_{\rm mig}.
\label{eq:hierarchy}
\ee
As mentioned in Sec.~\ref{sec:model}, the gaseous disk created by stellar mass-stripping will not align with the stellar orbit as long as the Lense-Thirring precession time is sufficiently rapid relative to the timescale for the disk to build-up mass near the minimum of the cycle, $\tau_{\Sigma}(\Sigma_{\rm min}) \sim \tau_{\rm cyc}/10$ (Eq.~\eqref{eq:tausigma}), a condition that is satisfied for large QPE periods, low SMBH masses, and/or high SMBH spin (see also Sec.~\ref{sec:demographics}). This is because only the component of the angular momentum added by the star which is parallel to the black hole spin axis will survive the precession-averaging process (see Fig.~\ref{fig:schematic}). 

Stated mathematically, if the disk is aligned with the black hole spin along the $\hat{z}$ direction, then the change in the total angular momentum of the disk $\vec{J}_{\rm disk}$ due to the addition of the stripped mass carrying the specific angular momentum of the stellar orbit $\vec{j}_{\star}$ roughly obeys,
\be
\frac{d \vec{J}_{\rm disk}}{dt} \simeq \langle \vec{j}_{\star} \rangle |\dot{M}_{\star}| \underset{\tau_{\rm prec} \ll \tau_{\rm cyc}/10}\simeq j_{\star,z}|\dot{M}_{\star}| \hat{z} = j_{\star}\cos \chi |\dot{M}_{\star}|\hat{z},
\ee
since the components of $\vec{j}_{\star}$ perpendicular to the black hole spin axis (e.g., $\hat{x}, \hat{y}$) cancel over multiple precession cycles (where $\langle... \rangle$ denotes a time-average). The order-unity factor $\cos \chi < 1$ implies that the stripped gas will ``circularize'' at a radius $r_{\rm c} \simeq r_0 \cos^{2}\chi$, slightly smaller than the collision radius $r_0$; however, outwards viscous spreading of the  mass deposited near $r_{\rm c}$ (e.g., \citealt{Metzger+12b}) will result in broadly similar disk properties as at $r_0$.

\subsection{Eccentric orbits}

Stars migrate to orbits of interest (i.e., those resulting in QPE flares, or generally interacting with TDE disks) primarily through gravitational-wave circularization from highly eccentric orbits of semi-major axis $a\gg r_{\rm p}$  (not to be confused with the SMBH spin parameter $a_\bullet$). Throughout the GW-dominated circularization phase, the stellar pericenter distance remains nearly constant at $r_{\rm p} \approx r_{\rm p,0}$. Thus, no significant star-disk interaction occurs during the GW inspiral phase, up until the orbit is mostly circularized, provided that $r_{\rm p,0} \gg r_{\rm T}$, where $r_{\rm T}$ is the tidal radii of the coincident TDEs progenitors. In this regime, the remaining eccentricity when TDE-triggered star-disk interaction first occurs is modest ($e \lesssim 0.5$), and our previous assumptions and estimates are generally valid.

If, however, $r_{\rm p,0} \gtrsim r_{\rm T}$, star-disk interaction may first ensue while the orbit is still highly eccentric. This scenario may also transpire if an AGN disk is present, extending to radii much greater than the size of a typical TDE disk \citep[e.g.,][]{Linial&Quataert24}. While the radius of the collision site between the star and the disk varies due to apsidal precession, the average (over the precession cycle) star-disk collision occurs at around $2 r_{\rm p}$ \citep{Linial&Quataert24}, with collision properties similar to those of a circular EMRI at radius $r_0 \approx r_{\rm p}$ and $P_{\rm QPE} \approx (P_{\rm orb}/2) (r_{\rm p}/a)^{3/2}$. Quantities that depend on the collision conditions, for example, $\Delta M_\star$, are essentially equivalent to the circular case, up to order unity corrections.

Interestingly, the typical timescale hierarchy diskussed in \S \ref{sec:orbit} is not necessarily maintained for highly eccentric orbits. For example, the stellar destruction time $\tau_{\rm dest} \approx (M_\star/\Delta M_\star) P_{\rm orb}$ may be \textit{longer} than the drag-induced circularization timescale
\begin{equation}
    \tau_{\rm mig} \equiv \tau_{\rm circ} \approx \frac{a}{|\dot{a}|} \approx \frac{M_\star}{\Delta M_{\rm coll}} \pfrac{r_{\rm p}}{a} P_{\rm orb} \,.
\end{equation}
with their ratio scaling as
\begin{equation}
    \frac{\tau_{\rm circ}}{\tau_{\rm dest}} \approx \alpha^{1/3} \pfrac{\eta R_\star \MBH}{r_{\rm p} M_\star}^{2/3} \pfrac{r_{\rm p}}{a}^{1/2} \,.
\end{equation}
The reversal of timescale hierarchy as $a \to r_{\rm p}$ suggests that the stellar destruction and circularization rates become comparable $\tau_{\rm dest} \approx \tau_{\rm circ}$ at some critical eccentricity, of roughly
\begin{equation}
    \left. 1-e \right|_{\tau_{\rm circ}=\tau_{\rm dest}} \approx 0.2 \; \pfrac{r_{\rm p}}{r_{\rm T}}^{4/3} \eta_{-3}^{-4/3} \alpha_{-1}^{2/3} M_{\bullet,6}^{-8/9} \mathcal{M}_\star^{4/3} \,,
\end{equation}
suggesting mild residual eccentricity when the stellar ablation exceeds the circularization rate (as $r_p \gtrsim r_{\rm T}$).

\begin{deluxetable}{lllll}
\tablecaption{Key Timescales ($\mathcal{M}_{\star} = \mathcal{R}_{\star} = 1$; $\alpha = 0.1$; $\eta = 10^{-3}$; $\xi_{\rm v} = 1$) \label{tab:timescales}}
\tablewidth{700pt}
\tabletypesize{\scriptsize}
\tablehead{
\colhead{Symbol} & $\beta$ & \colhead{Description} & 
\colhead{Value} & \colhead{Equation}
} 
\startdata
$P_{\rm QPE}$ & - & QPE period & $4\,{\rm hr}\; \mathcal{P}_{\rm QPE,4} \gtrsim P_{\rm QPE,min}$ & Eq.~\eqref{eq:PQPEmin} \\
$\tau_{\rm prec}$ & - & Nodal precession period & $1.4\,{\rm yr} \; \mathcal{P}_{\rm QPE,4}^{2}M_{\bullet,6}(a_\bullet/0.3)^{-1}$ & Eq.~\eqref{eq:Tprec} \\
$\tau_{\rm cyc}$ & $\beta = 0$  & Limit-cycle period & $7\,{\rm yr}\,\;\mathcal{P}_{\rm QPE,4}^{3.88}$ & Eq.~\eqref{eq:taucyc} \\
... & $0 \ll \beta < 3/16$  & ... & $7\,{\rm yr}\,\;(\beta/0.5)^{1/3} \mathcal{P}_{\rm QPE,4}^{2.78} M_{\bullet,6}^{2/9}$ & Eq.~\eqref{eq:taucyc2} \\
$\tau_{\nu}$ & $\beta > 3/16$ &  Viscous time & $4\,{\rm yr}\,\;\mathcal{P}_{\rm QPE,4}^{2.78}M_{\bullet,6}^{2/9}$ & Eq.~\eqref{eq:tvisc} \\
$\tau_{\rm dest}$& $\beta = 0$ & EMRI destruction time &  $100\,{\rm yr}\,\;\mathcal{P}_{\rm QPE,4}^{2.46}M_{\bullet,6}^{-2/3}$ & Eq.~\eqref{eq:taudest} 
\\
... & $0 \ll \beta < 3/16$ & ...  & $310\,{\rm yr}\, (\beta/0.1)^{2/3}\mathcal{P}_{\rm QPE,4}^{2/9}M_{\bullet,6}^{-2/9}$ & Eq.~\eqref{eq:taudest2} \\
... & $\beta > 3/16$ & ...  & $300\,{\rm yr}\,\;(1 + \frac{8}{3}\beta )\mathcal{P}_{\rm QPE,4}^{2/9}M_{\bullet,6}^{-2/9}$  & Eq.~\eqref{eq:taudestST} \\
$\tau_{\rm grav}$ & - & Gravitational torquing & $500\,{\rm yr} \; \mathcal{P}_{\rm QPE,4}M_{\bullet,6}$ & Eq.~\eqref{eq:taugrav} \\
$\tau_{\rm mig}$ & $\beta = 0$ & Drag-induced orbital migration & $6\times 10^{3}\,{\rm yr}\,\mathcal{P}_{\rm QPE,4}^{0.92}$ & Eq.~\eqref{eq:taumig} \\
... &  $0 \ll \beta < 3/16$ & ...  & $6\times 10^{4}\,{\rm yr}\, (\beta/0.1)^{2/3}\mathcal{P}_{\rm QPE,4}^{-1.32}M_{\bullet,6}^{4/9}$ & ... \\
... & $\beta > 3/16$ & ... &  $2.7\times 10^{3}\,{\rm yr}\,\;(1 + \frac{8}{3}\beta )\mathcal{P}_{\rm QPE,4}^{-2/9}M_{\bullet,6}^{2/9}$  & Eq.~\eqref{eq:taumig_ST}
\enddata
\end{deluxetable}

\subsection{Secular Evolution of QPE Quiescent Emission and Eruptions}
\label{sec:flares}

The time-averaged accretion rates predicted by our model $\dot{M}_{\rm acc} \gtrsim 0.1\dot{M}_{\rm Edd}$ broadly agree with those of observed QPE quiescent sources. However, depending on the disk/collision radius and the (uncertain) value of $\beta$, our calculations reveal that these rates can be achieved through a wide diversity of time-evolution behavior$-$ranging from limit cycles to quasi-steady accretion. It is furthermore not realistic to expect the one-zone model we have presented to capture the entire radially-dependent disk structure, which determines for example how faithfully the mass-inflow rate near the collision annulus reflects that reaching the central SMBH and setting the quiescent X-ray emission. Nevertheless, qualitative features of the black-hole accretion-rate evolution, such as characteristic timescales and the potential for both rising and falling light curves, are likely to be more robust. 

For thermally unstable evolution ($\beta < 3/16$, small $r_0 < r_{\rm rad}$), significant changes in the disk accretion rate occur on the limit-cycle duration $\tau_{\rm cyc}$ (Eq.~\eqref{eq:taucyc}, \eqref{eq:taucyc2}). The latter are typically years to decades, depending most sensitively on the mass-stripping efficiency $\eta$, the stellar radius, and the QPE period. Even for thermally-stable accretion ($\beta > 3/16$, large $r_0 > r_{\rm rad}$), gradual long-term variability in the mass accretion rate may still occur (e.g. due to viscous instabilities; \citealt{Lightman&Eardley74}), on the viscous timescale (Eq.~\eqref{eq:tvisc}) which is also years to decades. As with the average accretion rates, these diversity of timescales match those observed in QPE sources. The quiescent light curves in GSN 069 \citep{Miniutti+19,Miniutti+23b}, XMMSL1 J024916.6-041244 \citep{Chakraborty+21}, eRO-QPE3, eRO-QPE4 \citep{Arcodia+24} exhibit significant (factor of $\gtrsim \rm few$) changes on timescales from years to decades, including both rising and decaying emission (or phases of both in GSN 069; \citealt{Miniutti+23}), broadly consistent with our thermally-unstable accretion solutions.  The peak in the mass-accretion rate during the final ``flushing'' stage of each cycle evolves even faster in our models (on a timescale as short as weeks). In contrast, the apparently constant quiescent X-ray flux of RXJ 1301.9+2747 \citep{Giustini+20} over 20 years, and of eRO-QPE2 for $\sim 3.5$ years (Arcodia et al., in preparation), may point to collisionally-stabilized disk evolution in this system.  

Insofar that QPE eruptions are generated by the hot expanding debris clouds created by disk-star collisions \citepalias{Linial&Metzger23}, the eruption properties themselves would be expected to co-evolve along with the quiescent disk. This is indeed supported by observations which show secular changes in the QPE properties over multiple epochs in parallel with changes in the quiescent emission \citep{Miniutti+23b,Chakraborty+24,Arcodia+24,Pasham+24a}. 
\citetalias{Linial&Metzger23} estimate the duration and luminosities of the eruptions in terms of the disk properties near the collision annulus:
\be
t_{\rm QPE} \approx \left(\frac{\Delta M_{\rm coll}\kappa_{\rm T}}{4 \pi c v_{\rm k}}\right)^{1/2} \approx 0.13\,{\rm hr}\,(1 + \frac{8}{3}\beta )^{-1/2}(\eta_{-3}\xi_{\rm v}^{2})^{-1/6}\alpha_{-1}^{-1/3}\mathcal{M}_{\star}^{1/6}\mathcal{R}_{\star}^{1/3}\mathcal{P}_{\rm QPE,4}^{7/9}M_{\bullet,6}^{-5/18};
\label{eq:tQPE}
\ee
\be
L_{\rm QPE} \approx \frac{E_{\rm ej}(R_{\star}^{2}h)^{1/3}}{3 v_{\rm k}t_{\rm QPE}^{2}} \approx \frac{1}{6}L_{\rm Edd}\frac{(R_{\star}^{2}h)^{1/3}}{r_0} \approx 1.6\times 10^{41}\,{\rm erg\,s^{-1}}\left(\frac{\eta_{-3}\xi_{\rm v}^{2}}{\alpha_{-1}}\right)^{1/9}\frac{\mathcal{R}_{\star}^{10/9}}{\mathcal{M}_{\star}^{1/9}}\mathcal{P}_{\rm QPE,4}^{-20/27}M_{\bullet,6}^{20/27}.
\label{eq:LQPE}
\ee
Here, $\Delta M_{\rm coll} \simeq 2\pi R_{\star}^{2}\Sigma$ (Eq.~\eqref{eq:Mcoll}) and $E_{\rm ej} \simeq \Delta M_{\rm coll}v_{\rm k}^{2}/2$ are the mass and thermal energy, respectively, of disk material shocked by the star in a roughly head-on collision ($\chi \sim 1$), which expands above and below the midplane, powering the observed eruptions. In the final equality we have made use of our analytic estimates for the scale-height and surface density of thermally-stable disks (Eqs.~\eqref{eq:hoverr_stable},\eqref{eq:sigmastable}).

The bottom panel of Fig.~\ref{fig:beta0_limit} shows how these eruption properties evolve in time for the fiducial $\beta = 0$ model. The eruption luminosities $L_{\rm QPE}$ and duty cycles $t_{\rm QPE}/P_{\rm QPE}$ vary by almost an order of magnitude over a single cycle, illustrating how changes in the quiescent disk properties (locally, at $r_0$) could well manifest through secular evolution of the observed eruption properties. The cycle time-averaged values $\langle L_{\rm QPE} \rangle \sim 10^{41}$ erg s$^{-1}$, $\langle t_{\rm QPE}/P_{\rm QPE} \rangle \sim 10^{-2}$ are each roughly an order of magnitude lower than those of QPE eruptions ($L_{\rm QPE} \sim 10^{42}$ erg s$^{-1}$; $t_{\rm QPE}/P_{\rm QPE} \sim 0.1$). This may point to a larger effective stellar radius (for purposes of interacting with the disk) or a more massive SMBH than assumed in the fiducial model, or inaccuracies in the analytic estimates. 

Several effects may lead to an increase in the effective star-disk cross section producing the QPE flares. The star's outer layers may become inflated due to additional heating that may occur from the fraction of the collisional heating rate $\Delta M_{\rm coll} v_{\rm rel}^2 / P_{\rm QPE}$ deposited inside the star. The flare timing pattern observed in some of the known QPE systems indicate mild orbital eccentricities $e\approx 0.1$, suggesting significant tidal heating, possibly inflating the star's radius to about twice its unperturbed, main-sequence size. Repeated ablation of the star's outer layers may also indirectly increase the effective collision area. The bulk of the stripped material, $\Delta M_\star$, escapes the star at velocities comparable to $v_{\rm esc,\star} \approx \sqrt{GM_\star/R_\star}$. During the time between two consecutive collisions, the ablated material propagates a distance of order $v_{\rm esc,\star} P_{\rm QPE} \approx R_\star$, where we assumed $P_{\rm QPE} \gtrsim P_{\rm QPE,min}$. In an average sense, the star is thus shrouded by a substantial layer of ablated material, which itself can intersect the disk. Given that typically $\Delta M_{\star} \gg \Delta M_{\rm coll}$ and that the extent of the bulk of stripped material only doubles in radius, the effective surface density of the impacting shroud exceeds the disk's $\Sigma$. Hence, the ablated material is not expected to be substantially decelerated and incorporated into the disk in just a single collision, but rather over multiple orbital periods. This suggests that a complex flow composed of liberated disk material, ablated stellar debris, and possibly tidally stripped stellar material all evolve under the combined gravity of the SMBH and the star itself, governing the timing and emission properties of the resulting flares.  In this case, the effective stellar radius may not be limited by tidal gravity to the Roche lobe size, $R_{\rm RL} \approx 0.5 \, r_0 (M_\star/\MBH)^{1/3} \approx 1.1 \, R_\odot \, \mathcal{P}_{\rm QPE,4}^{2/3} \mathcal{M}_\star^{1/3}$. 

As discussed in Sec.~\ref{sec:orbit}, evolution of the EMRI's orbital semi-major axis or eccentricity is expected to be very slow (e.g., $\tau_{\rm mig} \gtrsim 10^{3}$ yr). Even in the presence of rapid nodal precession, the timing of disk collisions and hence the pattern of the twice-per-orbit periodic eruptions should be relatively steady over timescales of years if the accretion disk remains flat with its midplane aligned with the black hole spin axis (e.g., Fig.~\ref{fig:schematic}). However, if such alignment cannot occur or is incomplete, then the disk (or just the collision annulus) will also precess about the black hole spin axis, on a characteristic timescale comparable or shorter than $\tau_{\rm prec}$ (Eq.~\eqref{eq:Tprec}). This would give rise to a more complex temporal spacing of the eruptions (e.g., \citealt{Franchini+23}), which may be required to explain those QPE sources which exhibit less regular timing properties.  Indeed, {\it temporary} disk-spin alignment or warping could occur as substantial spin-misaligned angular momentum is added to the disk by stellar mass-ablation at the beginning of a limit-cycle (when $\Sigma$ is minimal), on a timescale shorter than the precession period (see also \citealt{XiangGruess+16}). We speculate this could explain why the repetition pattern of the eruptions from GSN 069, after their reappearance following the second quiescent-outburst, temporarily differed (it remains to be seen how well the repetition pattern will settled back into that observed prior to the outburst; G.~Minutti, private communication). 

A key property that enables QPEs to be detected is the higher temperatures of the eruptions $k_{\rm B}T_{\rm QPE} \sim 100-200$ eV \citep{Miniutti+19,Giustini+20,Arcodia+21,Chakraborty+21,Arcodia+22,Miniutti+23,Webbe&Young23} relative to those of the quiescent disk ($k_{\rm B}T \lesssim 50$ eV). \citetalias{Linial&Metzger23} found that the value of $k_{\rm B}T_{\rm QPE}$ radiated by the debris cloud created from a disk-star collision is sensitive to the disk's midplane density, insofar the rate of photon production within the hot shocked disk material scales as the density squared. The fact that our thermally-unstable models reveal that $\Sigma$ can vary by up to several orders of magnitude over its evolution (Figs.~\ref{fig:beta0_limit}, \ref{fig:loops}) therefore also implies variability in the eruptions temperatures$-$and hence in the detectability of QPE activity$-$over timescales of years to decades. This may lead to windows of time where QPE eruptions temporarily ``disappear'', if local thermal-instabilities indeed manifest. Such behavior is consistent with the cessation of QPE eruptions in GSN069 around the time of the second quiescent outburst \citep{Miniutti+23}. However, even for the most extreme limit cycles present in the $\beta = 0$ case, the disk spends over half of its time within a factor of a few of $\Sigma_{\rm max}$ (right panel of Fig.~\ref{fig:beta0_limit}).

\subsection{Accretion Activity in Galactic Nuclei and QPE Demographics}
\label{sec:demographics}

From Eq.~\eqref{eq:MdotTDE}, we see that fall-back accretion after a TDE remains above a given Eddington ratio $\dot{M}_{\rm TDE}/\dot{M}_{\rm Edd}$ for a timescale:
\be
t_{\rm Edd} \simeq t_{\rm TDE}\left(\frac{M_{\rm \star,TDE}}{3t_{\rm TDE}\dot{M}_{\rm TDE}}\right)^{3/5} \approx 6.8\,{\rm yr}\,\left(\frac{t_{\rm TDE}}{30\,{\rm d}}\right)^{2/5}\left(\frac{M_{\rm \star,TDE}}{M_{\odot}}\right)^{3/5}\left(\frac{\dot{M}_{\rm TDE}}{0.1\dot{M}_{\rm Edd}}\right)^{-3/5}.
\ee
By contrast, we have shown that a disk fed by a stellar-EMRI can maintain an accretion rate $\dot{M}_{\rm acc} \sim 0.1\dot{M}_{\rm Edd}$ (Eqs.~\eqref{eq:Mdotavg},\eqref{eq:MdotTS}) for a timescale $\tau_{\rm dest} \sim 10^{2}-10^{3}$ yr (Eqs.~\eqref{eq:taudest},\eqref{eq:taudest2},\eqref{eq:taudestST}), i.e. a factor of $\sim 10-10^{2}$ times longer than $t_{\rm Edd}$. Thus, even if stellar EMRIs occur $\sim 100$ times less frequently than TDEs in galactic nuclei (e.g., \citealt{Linial&Sari23}), their contribution to accretion activity in otherwise inactive galactic nuclei (i.e., those without large-scale AGN) can be comparable to that of TDEs.\footnote{The tidal disruption of giant stars has been argued to contribute $\sim 10\%$ of the total TDE rate  \citep{MacLeod+12} and to last for tens of hundreds of years, comparable to the disk lifetimes estimated in this work.  However, the susceptibility of their accretion streams to disruption by dense gas in the nucleus \citep{Bonnerot+16b} remains an uncertainty.}  However, unlike a TDE, such a source of ionizing UV/X-ray radiation lasting hundreds or thousands of years, will create a more spatially-extended narrow-line region in the nuclei of galaxies hosting interacting star/disk systems. This may be compatible with the estimated long duration of the accretion activity in QPE sources (e.g., \citealt{Wevers+23,Patra+23}). 

The per galaxy abundance of QPE sources has been recently constrained by \cite{Arcodia+24b} to be of order $\sim 10^{-5} \, \rm gal^{-1}$, which corresponds to a volumetric rate of $0.6\times 10^{-7} (\tau_{\rm dest}/10 \, \rm yr)^{-1} \; Mpc^{-3} \, yr^{-1}$. Thus, the relatively long QPE source lifetimes found by our calculations $\tau_{\rm dest} \sim 10^{2}-10^{3}$ yr  imply that a lower formation rate of QPE sources is needed to explain their observed occurrence rate in galactic nuclei than if the collision target is a comparatively short-lived TDE disk. This is relevant if rare conditions, such as a stellar EMRI with a mass/radius larger than the Sun (and correspondingly larger cross sections for interaction with the disk), or low inclination angles $\chi \ll 1$, are needed to produce detectable eruptions. However, such long-lived self-sustained disks obviously cannot be produced the disk-colliding body is a compact object, such as an intermediate-mass black holes (IMBH), instead of a nondegenerate star. While IMBH-disk collisions can generate QPE-like eruptions with similar properties to star-disk collisions \citepalias{Linial&Metzger23}, they were already disfavored as QPE sources because of their comparative rarity to stars and constraints based on the implied SMBH growth rate via IMBH cannibalism (\citetalias{Linial&Metzger23}). The inability of black holes to create self-sustained disks with which to interact, would only further reduce their appearance in the QPE population.

Although the sample size is still small, QPE sources appear to exhibit a preference for occurring in low-mass galaxies (e.g., \citealt{Wevers+22}) hosting correspondingly low-mass black holes, compared even to TDE host galaxies. This observational finding does not have a clear dynamical explanation in stellar-EMRI QPE models insofar that the TDE and EMRI rate should scale similarly with SMBH mass (however, see \citealt{Lu&Quataert23,King23}; \citetalias{Linial&Metzger23}). However, insofar that the presence of a gaseous disk is a necessary condition for QPE activity, our disk evolution models may offer new insights into QPE demographics. Firstly, the predicted EMRI destruction time scales inversely with the SMBH mass (e.g., $\tau_{\rm dest} \propto M_{\bullet,6}^{-2/3}$ for $\beta = 0$; Eq.~\eqref{eq:taudest}) such that QPE sources resulting from disk interaction will be longer-lived and hence more abundant at any given time in galactic nuclei which contain lower-mass SMBH. However, this effect may be in part compensated in a flux-limited sample if the eruptions are indeed more luminous from stars orbiting higher SMBH masses (see Eq.~\eqref{eq:LQPE}). A second reason that QPE sources might preferentially occur around low-mass SMBH relates to the requirement of a short precession time to maintain a large disk-star misalignment angle and hence a long-lived stellar-collision-fed disk (Fig.~\ref{fig:schematic}). This condition is more challenging to satisfy for larger black hole masses because $\tau_{\rm prec} \propto M_{\bullet}$. 

\section{Summary/Conclusions}
\label{sec:conclusions}

Motivated by questions about the origins and lifetimes of the black hole accretion flows in QPE systems, we have developed a model for the coupled evolution of a stellar EMRI on an inclined quasi-circular orbit
and the gaseous disk (modeled as a single annulus centered near their interaction radius). The model accounts for mass stripped from the star and thermal energy injected into the disk as the result of periodic disk-star collisions. 

Our main conclusions can be summarized as follows:
\begin{itemize}
\item While a transient gaseous disk (e.g., from an independent TDE) is necessary to initiate the process, stellar stripping soon becomes the dominant source of mass feeding the disk and the subsequent evolution ``forgets'' about the initial state. Qualitative features of the disk's evolution$-$particularly the presence or absence of limit-cycles$-$depend on the uncertain efficiency with which the mass stripped from the star heats the disk, as encapsulated within our model by the dimensionless parameter $\beta \simeq \dot{E}_{\rm abl}/|\dot{M}_{\star}|v_{\rm k}^2$. Future radiation hydrodynamic simulations of disk-star collisions are necessary to estimate the value of $\beta$ as a function of the system parameters such as the disk/star inclination angle.

\item For inefficient collision/ablation heating ($\beta < 3/16$) under radiation-dominated conditions, the disk is susceptible to the standard thermal instability. The instability manifests as limit-cycle behavior, in which the disk starts at low surface density $\Sigma \sim \Sigma_{\rm min}$ in the gas-pressure dominated radiatively-cooled ``low'' state, where it builds up mass gradually from stellar stripping. Eventually, once the surface density reaches a critical value $\Sigma \sim \Sigma_{\rm max}$, the disk transitions on the thermal time to an advection-dominated ``high'' state (Figs.~\ref{fig:beta0_evo},\ref{fig:beta0_limit}).  In the geometrically-thick high state, the disk rapidly ``flushes" until the high branches vanishes at $\Sigma < \Sigma_{\rm min}$ and the cycle starts again after the disk quickly cools back to the low branch. The total duration of the cycle is of order years to decades (Eq.~\eqref{eq:taucyc}, \eqref{eq:taucyc2}). 

\item As $\beta$ increases, collisional/ablation heating becomes relevant in offsetting radiative cooling in the low state and the boundaries of the limit cycles shrink (Fig.~\ref{fig:loops}; Appendix \ref{sec:analytics_intermediate_beta}). For sufficiently large $\beta$ above a critical threshold $\beta_{\rm stable} = 3/16$ (Appendix \ref{sec:stability}), collision/ablation heating is sufficient to completely stabilize the disk's thermal evolution, even under radiation-dominated conditions, resulting in a unique surface density, scale-height and mass accretion rate (Sec.~\ref{sec:stable}). 

\item Regardless of the value of $\beta \lesssim 1$, the time-averaged mass-accretion rates are characteristically $\langle \dot{M}_{\rm acc} \rangle \sim 0.03-0.3 \dot{M}_{\rm Edd}$ for QPE-like systems (Eqs.~\eqref{eq:Mdotavg}, \eqref{eq:MdotTS}). These rates, and the characteristic timescales over which they are predicted to vary due to changes in the disk properties at the collision radius ($\tau_{\rm cyc}$ or $\tau_{\nu}$), broadly agree with the observed quiescent emission of QPE sources. In particular, it is natural to expect both rising and decaying quiescent X-ray light curves over timescales of years to decades, e.g. without the need to invoke multiple full- or partial-TDEs in the same galactic nucleus.

\item In disk-star collision models for QPEs, time-dependence in the disk properties at the collision radius should naturally translate into secular timescale changes in the eruption properties, such as their luminosity, duration, and temperature (Fig.~\ref{fig:beta0_limit}).  For thermally-unstable disks, modulation of the eruption temperatures (relative to the comparatively stable quiescent disk temperature) due to large fluctuations in the disk midplane density (which controls photon production in the collision ejecta), could cause QPE sources to appear or disappear from detectability on timescales of years to decades. A reduction in the QPE temperature due to the higher disk density at the peak of the cycle may explain the temporary disappearance of detectable eruptions from GSN069 during its quiescent 2019 outburst \citep{Miniutti+19,Miniutti+23}, before their recent reappearance \citep{Miniutti+23b}.

\item Table \ref{tab:timescales} and Fig.~\ref{fig:beta_P_QPE_contours} summarize the key timescales of the problem. For the observed range of QPE periods $P_{\rm QPE} \sim 2-20$ hr, the star is completely ablated by disk collisions on a timescale of $\tau_{\rm dest} \sim 10^{2}-10^{3}$ yr, depending weakly on the free parameters of the problem (Eqs.~\eqref{eq:taudest2},\eqref{eq:taudestST}). However, the stellar lifetime may depend more sensitively on our simplified $\alpha-$viscosity disk model, which may over-estimate the midplane density (and hence under-estimate $\tau_{\rm dest}$) compared to radiation GRMHD simulations (e.g., \citealt{Jiang+19}). The destruction time also depends on our assumed prescription for stellar mass ablation (Eq.~\eqref{eq:Mdotstar}), which though supported by numerical simulations performed in the regime $h \sim R_{\star}$, requires further elucidation through targeted studies (e.g., Yao \& Quataert, in prep), including both the $h \ll R_{\star}$ and $h \gg R_{\star}$ limits. Across most of the parameter space, the star is destroyed before its orbit will experience significant migration due to gas drag or gravitational interaction with the disk. 

\item Despite their comparatively rare formation rate (e.g., relative to TDEs), the longer durations stellar collision-fed disks spend accreting at a substantial fraction of the Eddington accretion rate may have a number of consequences. They should produce radially-extended narrow-line emission regions, consistent with those observed from QPE hosts \citep{Wevers+22,Patra+23}, in $\sim 0.1\%$ of low-mass galaxies (for an assumed EMRI rate of $10^{-6}$ yr$^{-1}$ per galaxy). Longer QPE source lifetimes alleviate constraints on the fraction of stellar EMRIs that need to produce detectable eruptions in order to reproduce the observed QPE source population. 

\item Our model assumes that the accretion disk remains on averaged aligned with the black hole spin axis, which is justified provided that the stellar orbit precesses sufficiently rapidly compared to the timescale over which the disk surface density evolves (Fig.~\ref{fig:schematic}). While this condition is trivially satisfied for thermally-stable disks with long viscous timescales $\tau_{\nu} \gg \tau_{\rm prec}$, for the limit-cycle case the more stringent condition $\tau_{\rm prec} \lesssim \tau_{\rm cyc}/10$ requires a combination of a low-mass and/or rapidly-spinning SMBH. Even temporary changes in the star-disk inclination angle resulting from the angular momentum of the deposited stripped mass (e.g., \citealt{XiangGruess+16}) could result in short-lived changes to the QPE eruption recurrence time pattern, similar that observed in GSN 069 following their recent reappearance. 

\end{itemize}

\begin{acknowledgments}
We thank Riccardo Arcodia, Margherita Giustini, Giovanni Miniutti, and Eliot Quataert for comments on the manuscript. IL acknowledges support from a Rothschild Fellowship and The Gruber Foundation.  BDM was supported, in part, by the National Science Foundation (grant No. AST-2009255). The Flatiron Institute is supported by the Simons Foundation.
\end{acknowledgments}

\appendix

\section{Analytic Estimates of Thermally Unstable Disk Evolution}
\label{sec:analytics_intermediate_beta}

As in the $\beta = 0$ case (Sec.~\ref{sec:beta0}), the results for finite collision/ablation heating ($0 \ll \beta < \beta_{\rm stable}$) can be understood analytically. Balancing now radiative cooling with both accretion {\it and} collision/ablation heating ($\dot{E}_{\rm abl} + \dot{E}_{\rm acc} \approx \dot{E}_{\rm rad}$) gives the following expression for the disk surface density:
\begin{equation}
    \Sigma_{\dot{E}_{\rm rad} = \dot{E}_{\rm abl} + \dot{E}_{\rm acc}} \approx \Tilde{A} T^{8/3} \left( 1 - \Tilde{B} T^{4} \right)^{1/3} \,,
    \label{eq:Sigcoll}
\end{equation}
where 
\begin{equation}
    \Tilde{A} = \left( \frac{\pi}{18 \beta \eta} \frac{c a^2}{\kappa} \pfrac{R_\star}{r_0}^{-4} \pfrac{\MBH}{M_\star}^{-1} \frac{r_0^2}{v_{\rm k}^5} \right)^{1/3}; \Tilde{B} = \frac{3 r_0 \alpha a \kappa }{8 v_{\rm k} c}.
\end{equation}
Here we have assumed $p_{\rm rad} \gg p_{\rm gas}$ and $\kappa \approx \kappa_{\rm es}$.  

As the disk surface density and temperature $T \propto \Sigma^{3/8}$ rise during each cycle, thermal stability is lost once the accretion heating term (first term on the right hand side of Eq.~\eqref{eq:Sigcoll}) becomes comparable to collision/ablation heating.  This occurs at a maximum density
\begin{multline}
    \Sigma_{\rm max} = \Tilde{A} \pfrac{4}{27 \Tilde{B}^2}^{1/3} =
    \frac{4}{9} \pfrac{2\pi}{3}^{1/3} \frac{1}{\kappa_{\rm es}} \pfrac{c}{v_{\rm k}} \left( \frac{1}{\beta \eta \alpha^2} \pfrac{r_0}{R_\star}^4 \pfrac{M_\star}{\MBH} \right)^{1/3} \approx \\
    1.9\times 10^4 \, {\rm g \, cm^{-2}} \; (\beta/0.1)^{-1/3} \eta_{-3}^{-1/3} \alpha_{-1}^{-2/3} \mathcal{M}_\star^{1/3} \mathcal{R}_\star^{-4/3} M_{\bullet,6}^{-2/9} \mathcal{P}_{\rm QPE,4}^{11/9} \,,
\label{eq:sigma_intermediate}
\end{multline}
as is attained at temperature $T = (2/(3\Tilde{B}))^{1/4}$.  The time spent at around $\Sigma \simeq \Sigma_{\rm max}/2$ each phase, is then approximately
\begin{equation}
    \tau_{\Sigma} \approx \frac{\pi r_0^2 \Sigma_{\rm max}}{|\dot{M}_\star|} \approx \frac{1}{2} P_{\rm QPE} \, \eta^{-1} \pfrac{r_0}{R_\star}^4 \pfrac{M_\star}{\MBH} \pfrac{h}{r} \,,
\end{equation}
with
\begin{multline}
    \frac{h}{r} = \frac{a T^4 r_0}{3 \Sigma v_{\rm k}^2} = \frac{2}{9} \frac{a r_0}{\Tilde{B} \Sigma v_{\rm k}^2} = 2 \pfrac{4}{9\pi}^{1/3} \left( \frac{\alpha}{\beta \eta} \pfrac{r_0}{R_\star}^4 \pfrac{M_\star}{\MBH} \right)^{-1/3} \approx \\
    8.8\times 10^{-3} \; (\beta/0.1)^{1/3} \eta_{-3}^{1/3} \alpha_{-1}^{-1/3} \mathcal{R}_\star^{4/3} \mathcal{M}_\star^{-1/3} \mathcal{P}_{\rm QPE,4}^{-8/9} M_{\bullet,6}^{-1/9} \,.
\end{multline}
The time spent at any surface density, $\Sigma<\Sigma_{\rm max}$, $\tau_{\Sigma} \propto h/r \propto \Sigma^{1/2}$ is again dominated by large $\Sigma$. The total cycle duration can thus be estimated as:
\begin{multline}
    \tau_{\rm cyc} \approx \tau_{\rm \Sigma_{\rm max}} \approx P_{\rm QPE} \, \eta^{-2/3} \, \pfrac{r_0}{R_\star}^{8/3} \pfrac{M_\star}{\MBH}^{2/3} \pfrac{\beta}{\alpha}^{1/3} \pfrac{4}{9\pi}^{1/3} \approx \\
    3.4 \, {\rm yr} \; \eta_{-3}^{-2/3} \alpha_{-1}^{-1/3} (\beta/0.1)^{1/3} \mathcal{M}_\star^{2/3} \mathcal{R}_\star^{-8/3} \mathcal{P}_{\rm QPE,4}^{25/9} M_{\bullet,6}^{2/9} \,.
    \label{eq:taucyc2}
\end{multline}
Similar to Eqs.~\eqref{eq:Mdotavg}, \eqref{eq:taudest}, the mass-averaged accretion rate and stellar destruction time are given by:
\be
\langle \dot{M}_{\rm acc} \rangle \approx \frac{\pi r_0^{2}\Sigma_{\rm max}}{\tau_{\rm cyc}} \approx 0.07 \, \dot{M}_{\rm Edd}  \; \eta_{-3}^{1/3} \alpha_{-1}^{-1/3}(\beta/0.1)^{-2/3}\mathcal{R}_{\star}^{4/3}\mathcal{M}_{\star}^{-1/3}\mathcal{P}_{\rm QPE,4}^{-2/9} M_{\bullet,6}^{-7/9}
\ee
\be
\tau_{\rm dest} \simeq \frac{9}{16}\frac{M_{\star,0}}{\langle \dot{M}_{\rm acc} \rangle} \approx 310 \,{\rm yr}\,\eta_{-3}^{-1/3} \alpha_{-1}^{1/3} (\beta/0.1)^{2/3}\mathcal{R}_{\star}^{-4/3}\mathcal{M}_{\star}^{4/3}\mathcal{P}_{\rm QPE,4}^{2/9}M_{\bullet,6}^{-2/9},
\label{eq:taudest2}
\ee
where the factor $\approx 9/16$ accounts for the dependence $\tau_{\rm dest} \propto \mathcal{M}_\star^{16/9}$ as the star loses mass adiabatically ($\mathcal{R}_{\star} \propto \mathcal{M}_{\star}^{-1/3}$; Eq.~\eqref{eq:Rstar}). Eq.~\eqref{eq:taudest2} depends weakly on most of the parameters, accounting also for the fact that lower-main sequence stars roughly obey $\mathcal{R}_{\star} \propto \mathcal{M}_{\star}$.

\section{Thermal Stability with Collision/Ablation Heating}
\label{sec:stability}

Neglecting radial advection, the disk cools through radiation, at the rate (Eq.~\eqref{eq:dEthdt})
\begin{equation}
    \dot{E}_- = \dot{E}_{\rm rad} \approx \frac{4\pi}{3} r_0^2 \sigma_{\rm SB} \frac{T^4}{\kappa \Sigma} = \frac{1}{4} L_{\rm Edd} \pfrac{h}{r}\,,
\end{equation}
where in the second equality we have assumed $p_{\rm rad} \gg p_{\rm gas}$ and $\kappa \approx \kappa_{\rm es}$. The total heating rate, including viscous accretion and collision/ablation heating, can be written (Eq.~\eqref{eq:dEthdt}):
\begin{eqnarray}
    \dot{E}_{+} = \dot{E}_{\rm acc} + \dot{E}_{\rm abl} = \left( \frac{3}{8}|\dot{M}_{\rm acc}| + \beta |\dot{M}_\star| \right) v_{\rm k}^2 \,.
\end{eqnarray}
We have neglected heating from the disk material directly shocked by the star because $|\dot{M}_{\star}| \gg \dot{M}_{\rm coll}$ in our fiducial models (Figs.~\ref{fig:beta0_evo}). 

Holding $\Sigma$ constant (since it varies on the longer, viscous timescale), the criterion for thermal     stability can be written (e.g., \citealt{Piran78}):
\begin{equation}
    \pfrac{\partial \log{\dot{E}_{+}}}{\partial h} < \pfrac{\partial \log{\dot{E}_{-}}}{\partial h} \,,
\end{equation}
or equivalently,
\begin{equation}
    \frac{2 \frac{3}{8} \dot{M}_{\rm acc}-\beta \dot{M}_{\star}}{\frac{3}{8} \dot{M}_{\rm acc} + \beta \dot{M}_{\star}} < 1 \,.
\end{equation}
In steady-state, $\dot{M}_{\rm acc} \approx \dot{M_\star}$, and thus stability requires $\beta > \beta_{\rm stable} = 3/16$.  This can be understood because (at fixed $\Sigma$), while accretion heating is unstable $\dot{E}_{\rm acc} \propto h^2$ (e.g., \citealt{Lightman&Eardley74}), collision/ablation heating $\dot{E}_{\star} \propto h^{-1}$ is stabilizing and hence can dominate for sufficiently large $\beta$.

\bibliographystyle{aasjournal}
\bibliography{QPEdiskevo}

\end{document}